\documentclass{article}
 \usepackage{graphicx}
\usepackage{amsmath}
\usepackage{epsfig}
\usepackage{epstopdf}

\usepackage{amsmath,amssymb,amsfonts,txfonts,graphics,psfrag}

\DeclareMathOperator{\OSp}{OSp}
\DeclareMathOperator{\SO}{SO}
\DeclareMathOperator{\SU}{SU}
\DeclareMathOperator{\OO}{O}
\DeclareMathOperator{\UU}{U}
\DeclareMathOperator{\SL}{SL}
\DeclareMathOperator{\ssp}{sp}
\DeclareMathOperator{\so}{so}
\DeclareMathOperator{\osp}{osp}
\DeclareMathOperator{\End}{End}
\DeclareMathOperator{\ssl}{sl}

\DeclareMathOperator{\sgn}{sgn}
\DeclareMathOperator{\sdim}{sdim}

\DeclareMathOperator{\Hom}{Hom}
\DeclareMathOperator{\tr}{tr}

\DeclareMathOperator{\str}{str}
\DeclareMathOperator{\sch}{sch}
\DeclareMathOperator{\dg}{g}
\DeclareMathOperator{\Sym}{Sym}
\DeclareMathOperator{\GL}{GL}
\DeclareMathOperator{\gl}{gl} 
\DeclareMathOperator{\strans}{st}
\DeclareMathOperator{\sdet}{sdet}
\DeclareMathOperator{\trans}{t}
\DeclareMathOperator{\const}{cst}

\DeclareMathOperator{\rad}{rad}

\makeatletter
\@addtoreset{equation}{section}
\makeatother

\def\bea{\begin{eqnarray}}
\def\eea{\end{eqnarray}}
\def\be{\begin{equation}}
\def\ee{\end{equation}}

\begin{document}

\def\wgta#1#2#3#4{\hbox{\rlap{\lower.35cm\hbox{$#1$}}
\hskip.2cm\rlap{\raise.25cm\hbox{$#2$}}
\rlap{\vrule width1.3cm height.4pt}
\hskip.55cm\rlap{\lower.6cm\hbox{\vrule width.4pt height1.2cm}}
\hskip.15cm
\rlap{\raise.25cm\hbox{$#3$}}\hskip.25cm\lower.35cm\hbox{$#4$}\hskip.6cm}}

\def\wgtb#1#2#3#4{\hbox{\rlap{\raise.25cm\hbox{$#2$}}
\hskip.2cm\rlap{\lower.35cm\hbox{$#1$}}
\rlap{\vrule width1.3cm height.4pt}
\hskip.55cm\rlap{\lower.6cm\hbox{\vrule width.4pt height1.2cm}}
\hskip.15cm
\rlap{\lower.35cm\hbox{$#4$}}\hskip.25cm\raise.25cm\hbox{$#3$}\hskip.6cm}}

\def\begeqar{\begin{eqnarray}}
\def\endeqar{\end{eqnarray}}

%---------------------------------------------------------------------
%
%
%-------------------------------------------------------------
%
%

\begin{center}

%%%%%%%%%%%%%%%%%%%%%%%%%%%%%%%%%%%%%%%%%%%%%%%%%%%%%%%%%%%%%%%%%%%%%%%%%%%%%%%

\Large{A lattice approach to  the  conformal $\OSp(2S+2|2S)$   supercoset sigma 
model}
\Large{Part I: Algebraic structures in the spin chain. The Brauer algebra.}
%%%%%%%%%%%%%%%%%%%%%%%%%%%%%%%%%%%%%%%%%%%%%%%%%%%%%%%%%%%%%%%%%%%%%%%%%%%%%%%

\vskip 1cm

Constantin Candu$^{(1)}$  and Hubert Saleur$^{(1,2)}$
\vspace{1.0em}

%% Address:
%%

{\sl\small  Service de Physique Th\'eorique, CEA Saclay,\\
Gif Sur Yvette, 91191, France$^{(1)}$\\}
{\sl\small  Department of Physics and Astronomy,
University of Southern California\\
Los Angeles, CA 90089, USA$^{(2)}$\\}

\end{center}

\begin{abstract}
We define and study  a lattice model which we argue is in the universality class
of the $\OSp(2S+2|2S)$   supercoset sigma model for a large range of values of
the coupling constant $g_\sigma^2$. In this first paper, we analyze in details
the symmetries of this lattice  model, in particular the decomposition of the
space of the quantum spin chain $V^{\otimes L}$ as a bimodule over
$\OSp(2S+2|2S)$ and its commutant, the Brauer algebra $B_L(2)$. It
turns out that $V^{\otimes L}$ is a nonsemisimple module for both 
$\OSp(2S+2|2S)$ and $B_L(2)$. The results are used in the companion paper to
elucidate the structure of the (boundary) conformal field theory.

\end{abstract}

\section{Introduction }

The solution of the   $AdS_5\times S^5$ worldsheet  string theory is one of the cornerstones of the 
AdS/CFT duality program. Despite continuous effort and progress on classical aspects in particular
\cite{Bena}, and the generally accepted presence of both integrability and
conformal invariance symmetries, most aspects of the quantum theory remain
elusive. 

It is natural to try to understand some aspects of this quantum theory by first tackling simpler models 
with similar properties. The so called $\OSp(2S+2|2S)$ coset model - specifically, a sigma model on the supersphere $\OSp(2S+2|2S)/\OSp(2S+1|2S)$ - is a very attractive candidate for such an exercise: like the $AdS_5\times S^5$ worldsheet theory it is conformal invariant and its target space is a supergroup coset. Of course, it lacks other aspects such as the BRST structure of the string theory. 

Apart from the string theory motivation, models such as the $\OSp(2S+2|2S)$ coset model are extremely interesting from the pure conformal field theory point of view. Indeed, they are sigma models which 
are massless without any kind of topological term, and for a large range of values of the coupling constant $g_\sigma^2$. To make things more precise let us briefly remind the reader 
of some generalities.  Supersphere sigma models
have target super space the supersphere
$S^{R-1,2S}:=\OSp(R|2S)/\OSp(R-1|2S)$ and can be viewed 
as a ``supersymmetric'' extension of the 
nonlinear $\OO(N)$ sigma models (which differs of course from 
the usual $\OO(N)$ ``supersymmetric''  models). Use as coordinates a real scalar
field
\begin{equation*}
  \phi:= (\phi^{1},\ldots,\phi^{R+2S})
\end{equation*}
where the first $R$ components are bosons, the last $2S$ ones 
fermions, and the invariant bilinear form
\begin{equation*}
 \phi.\phi'=\sum J_{ij}\phi^{i}\phi'^{j}
\end{equation*}
where $J$ is the orthosymplectic metric
\begin{equation*}
    J=\left(\begin{array}{ccc}
    I_{R}&0&0\\
    0&0&-I_{S}\\
    0&I_{S}&0\end{array}\right)
\end{equation*}
$I$ denoting the identity. The unit supersphere is defined by 
the constraint
\begin{equation*}
     \phi.\phi=1
\end{equation*}
The action of the sigma model (conventions are that the Boltzmann 
weight is $e^{-S}$) reads
\begin{equation*}
S={1\over 2g_{\sigma}^{2}}\int d^{2}x~
\partial_{\mu}\phi.\partial_{\mu}\phi
\end{equation*}
The perturbative $\beta$ function depends only on $R-2S$ to all 
orders (see, e.g., Ref \cite{Wegner}), and is the same as the one of the
$\OO(N)$ model with
$N:=R-2S$. Physics can be reliably understood from the first 
order beta function
\begin{equation*}
 \beta(g_{\sigma}^{2})=(R-2S-2)g_{\sigma}^{4}+O(g_{\sigma}^{6})
\end{equation*}
The model for $g_{\sigma}^{2}$ positive
flows to strong coupling for $R-2S>2$. 
Like in the ordinary sigma models case, the symmetry is restored at 
large length scales, and the field theory is massive. For $R-2S<2$ 
meanwhile, the model flows to weak coupling, and 
the symmetry is spontaneously  broken. One expects this scenario to 
work for $g_\sigma^2$ small enough, and the  corresponding
 Goldstone phase to be separated from a non perturbative 
strong coupling phase 
by a critical point.

The case we are interested in here is  $R-2S=2$, where the $\beta$ function 
vanishes to all orders in perturbation theory, and the model is 
expected to be conformal invariant, at least for $g_\sigma^2$ small enough, the
Goldstone phase being replaced by a phase with continuously varying
exponents not unlike the low temperature Kosterlitz Thouless phase.
How the group symmetry combines with the (logarithmic) conformal symmetry 
in such models is largely unknown. It is an essential question to be solved before any serious attempts to understanding universality classes in non interacting disordered 2D electronic systems can be contemplated \cite{Zirnbauer}.

The $\OSp(2S+2|2S)$ coset model was considered in particular in two papers by Mann and Polchinski using the massless scattering and Bethe ansatz approaches. This is indeed a natural idea, since 
supersphere sigma models are in general integrable, and, when
massive (ie $R-2S>2$) can be described by a scattering theory
involving particles in the fundamental representation of the group.
The $S$ matrix is well known
\begin{equation*}
\check{S}(\theta)=\sigma_{1}(\theta)E+\sigma_{2}(\theta)P+\sigma_{3}
(\theta)I
\label{smat}
\end{equation*}
Here, $I$ is the identity, $P$ is the graded permutation operator,
and $E$ is proportional to the projector on the identity
representation.
For $R,S$ arbitrary, factorizability requires that
\begin{eqnarray*}
  \sigma_{1}(\theta)&=&-{2i\pi\over
(N-2)(i\pi-\theta)}\sigma_{2}(\theta)\nonumber\\
  \sigma_{3}(\theta)&=&-{2i\pi\over
(N-2)\theta}\sigma_{2}(\theta)\label{genweights}
  \end{eqnarray*}
where $N=R-2S$, while $\sigma_{2}$ itself is determined, up to CDD 
factors, by
crossing symmetry and unitarity. One immediately observes that when
$N=2$, the amplitude $\sigma_{2}$ cancels out, leaving a scattering
matrix with a simpler tensorial structure, since the $P$ operator
disappears.  This corresponds  to a
particular point \cite{ReadSaleur01} on the sigma model critical line (where, among other
things,  the symmetry
is enhanced to $\SU(2S+2|2S)$), the rest of which
is not directly accessible by this construction.\footnote{The case
$R=2,S=0$ is special and allows for an extension of the S matrix to
the whole $\OO(2)$ critical line.} The idea used in 
\cite{MannPolchinski} is to consider an analytical continuation to $R,S$ real, and an
approach to $R-2S=2$ with proper scaling of the mass. Though interesting results were obtained, the emphasis in these papers was not on conformal properties.  

Another line of attack, more suited to the conformal aspects, was launched by Read and Saleur in 2001 \cite{ReadSaleur01}, who proposed to 
use a lattice regularization to control the integrable features of the model. They obtained in this way the spectrum of critical exponents for several related sigma models on super target spaces, including the  $\OSp(2S+2|2S)$ coset one at a particular (critical) value of the coupling $g_\sigma^2$. The results exhibited several mysterious features, including a pattern of large degeneracies, and a set of values of the exponents covering (modulo integers) all the rationals. In two subsequent papers \cite{ReadSaleur07i,ReadSaleur07ii}, it was argued further that many algebraic properties  of the conformal field theory could be obtained at the lattice level already. These include fusion, and the structure of conformal ``towers'' (see below for further details).

The work we present in this paper and its companion is an attempt at understanding the conformal field theoretic description of the $\OSp(2S+2|2S)$ model for all values of the coupling by using a lattice regularization. Foremost in the lattice approach is the understanding of the algebraic structure of the lattice model - the algebra defined by the local transfer matrices and its commutant. While in the cases
discussed in \cite{ReadSaleur07ii} most necessary results were already available in the mathematical literature, the situation here is much more complicated: in a few words, we have to deal, instead of the Temperley Lieb algebra, with the {\sl Brauer algebra} whose representation theory, in the non semi-simple case, is far from fully understood. An important part of our work has consisted in filling up the necessary gaps of the literature, sometimes rigorously, but sometimes at the price of some conjectures. This algebraic work is the subject of the first paper, which we realize might be a bit hard to read for a physics reader. We capitalize on the algebraic effort in the companion paper, where the boundary  conformal field theory  for the coset sigma model is analyzed thoroughly.

In the second section of this paper we discuss generalities about lattice
regularizations of $\OO(N)$ sigma models in 2 dimensions and define the model we
shall be interested in. In section 3, the transfer matrix, the loop
reformulation and the associated Brauer algebra are introduced and discussed.
Section 4 is the main section, where the full decomposition of the Hilbert space
of the lattice model under the action of $\OSp(2S+2|2S)$ and $B_L(2)$ is
obtained. Our main result can be found in eqs.~\eqref{eq:rez1} and 
\eqref{eq:rez2}. Section 5 discusses aspects of the hamiltonian limit and
section 6 contains conclusions. Technical aspects of representation theory are
discussed further in the appendices. 

For the reader's convenience, we provide here a list 
of notations used throughout the paper:
 \begin{itemize}
\item $\osp(R|2S)$ is the Lie superalgebra of the supergroup
  $\OSp(R|2S)$
\item $B_L(N)$ is the Brauer algebra on $L$ strings with fugacity for
  loops $N=R-2S$
\item $V_{R|2S}=\mathbb{C}^{R|2S}$ is the mod 2 graded vector space $\mathbb{C}^R \oplus
  \mathbb{C}^{2S}$ with even part $V_0=\mathbb{C}^R$ and odd part
  $V_1=\mathbb{C}^{2S}$. We shall often drop the indices $R,2S$ in $V_{R|2S}$.
\item $V^{\otimes L}$ is considered as a left $\osp(R|2S)$ and right  $B_L(N)$
  bimodule
\item $\lambda\vdash L$ stands for ``$\lambda$ is a partition of $L$''
  and $\lambda'$ is the partition $\lambda$ transposed
\item $\Sym(L)$ and $\mathbb{C}\Sym(L)$ are the symmetric group on $L$
  objects and its group algebra
\item $T_L(q)$ is the Temperley Lieb algebra with fugacity for loops $q+q^{-1}$
\item $d$ and $D$ are generic elements of $B_L(N)$ and $\OSp(R|2S)$
\item $X_L=\{\,\mu\vdash L-2k\mid k=0,\dots,[L/2]\,\}$ is the set of
  partitions labeling the  weights of $B_L(N)$.
  $X_L(S)\subset X_L$ selects
  those of them which do realize on $V^{\otimes L}$
\item Associate weights $\lambda,\lambda^*$ are labels of
  $\OSp(R|2S)$ irreps which are nonequivalent(identical) and
  become(split into two) isomorphic(nonequivalent) irreps under the restriction
to the proper subgroup $\OSp^+(R|2S)$ of supermatrices with $\sdet D =
  +1$
\item $H_L(S)=\{\,\lambda\in X_L(S)\mid \lambda_{r+1}\leq S\,\}$ is
  the set of  hook shape partitions labeling the weights of
  $\osp(R|2S)$ irreps appearing in $V^{\otimes L}$ and  $Y_L(S) =
  H_L(S)\cup H_L(S)^*$
\item $\Delta_L(\mu)$ are standard or generically irreducible representations of $B_L(N)$
\item $S(\lambda)$, $g(\lambda), G(\mu)$, $B_L(\mu)$, $D_L(j)$ are irreducible representations
  of $\mathbb{C}\Sym(L), \osp(R|2S)$, $\OSp(R|2S)$, $B_L(N)$ and
  $T_L(q)$ respectively.
\item $\mathcal{I}g(\lambda), \mathcal{I}G(\mu),  \mathcal{I}B_L(\mu)$
  are direct summands of $V^{\otimes L}$ as a $\osp(R|2S)$, $\OSp(R|2S)$
  and $B_L(N)$ module
\item $sc_\lambda$ are the supersymmetric generalization of $\OO(N)$
  symmetric functions
\item $\chi'_\mu(d), \chi_\mu(d), \sch_\lambda(D)$ are characters of
  $\Delta_L(\mu)$, $B_L(\mu)$ and $G(\lambda)$.
\end{itemize}

\section{The $\OSp(R|2S)$ lattice models: generalities}

\subsection{The models and their loop reformulations}

Lattice discretizations of $\OO(N)$ sigma models have had a long 
history. The simplest way to go is obviously to introduce spins 
taking 
values in the target manifold  - the sphere $\OO(N)/\OO(N-1)$  - on the 
sites of a discrete lattice, with an 
interaction energy of the Heisenberg type 
$E=-J\sum_{<ij >}\vec{S}_{i}.\vec{S}_{j}$ (where $.$ stands for the 
bilinear $\OO(N)$-invariant quadratic form). This is however 
difficult 
to study technically, as the number of degrees of freedom on each 
site 
is infinite. A possible way to go is to discretize the target space, 
leading to various types of ``cubic models'' \cite{Schick}.
%\footnote{\textbf{The fundamental and adjoint representation of 
%$\OO{(N)}$ will remain irreducible when restricted to
%the symmetry group of the $N$ dimensional hypercube --- the 
%hyperoctahedral group $W_N$.
%This may perhaps help to ask the question if  $W_N$ and $\OO{(N)}$ 
%symmetry are in the
%same universality class (the centraliser is the same).
%I mean if one can consider the spins $S_i$ to be Ising type, without 
%changing the universality class.}}.
Another way which has proved especially fruitful in two dimensions has
been to reformulate the problem of calculating the partition or 
correlation functions geometrically by using the techniques of high 
or low temperature expansions, thus obtaining graphs with complicated 
interaction rules and weights determined by properties of the 
underlying groups. 
The simplest of these formulations appeared in \cite{Domany} where the 
authors   studied 
the $\OO(N)$ model on the honeycomb lattice in two dimensions, and 
replaced moreover the 
term $\prod_{<ij>}e^{\beta J\vec{S}_{i}.\vec{S}_{j}}$ by its 
considerably simpler high temperature approximation 
$\prod_{<ij>}(1+K \vec{S}_{i}.\vec{S}_{j})$, $K=\beta J$  \cite{Nienhuis}. 
Expanding the brackets,
in say the calculation of the partition function, one can draw graphs 
by putting a bond between neighboring sites $i$ and $j$ whenever the 
term $\vec{S}_{i}.\vec{S}_{j}$ is picked up. The integral over spin 
variables leaves only loops, with a fugacity equal to $N$ as there 
are $N$ colors one can contract. Note that because of the very low 
coordination number of the honeycomb lattice, only self-avoiding 
loops are obtained. This leads to the well known self-avoiding loop gas partition 
function:
\begin{equation}
    Z_{SAL}=\sum_{{\cal G}}K^{E}N^{L}\label{nien_triv}
    \end{equation}
where the sum is taken over all configurations $\mathcal{G}$ of self avoiding,
mutually avoiding closed loops in number $L$, covering a total of $E$
edges.
Note that once an expression such as (\ref{nien_triv}) is written 
down, it is possible to analytically continue the definition of the 
model for $N$ an arbitrary real number. Barring the use of 
superalgebras, only $N$ integer greater or 
equal to one has a well defined meaning as a spin model (the case 
$N=1$ coincides with the Ising model~\footnote{The group $\OO(2)$ is 
different from 
$\SO(2)\simeq \UU(1)$ 
because of the additional $\mathbb{Z}_{2}$ freedom in choosing the sign of the 
determinant.}). In two 
dimensions, the Mermin Wagner theorem prevents spontaneous symmetry 
breaking, so for $N$ integer, critical behavior can only occur for 
$N=1,2$. Analysis of the 
same beta function suggests however 
that  lattice models defined by suitable analytic continuation should 
have  a Goldstone low  temperature phase 
for all $N<2$, though it says nothing about whether  this phase might 
end by a second or first order phase transition.

Model (\ref{nien_triv}) lacks interaction  terms which 
would appear with less drastic choices of the lattice and the 
interactions: these are the terms where the loops intersect, either 
by going over 
the same edge, or over the same vertex, maybe many times. It has 
often been argued that 
such terms are irrelevant for the study of the critical points of 
the $\OO(N)$ models in two dimensions. Most of the 
interest has focused on such critical points for $N\in [-2,2]$, 
which 
have geometrical applications - in particular the case $N=0$ is 
related with the physics of self-avoiding walks.  It  turns out 
however that intersection terms are crucial for the understanding of 
low temperature phases. Indeed, the model (\ref{nien_triv}) {\sl does 
have} a sort 
of Goldstone phase for $N\in [-2,2]$ called the dense phase, but its 
properties are not 
generic, and destroyed by the introduction of a small amount of 
intersections. A simple way to 
see that the dense  
 phase is not generic is  that the exponents at $N=2$ are always 
 those of the Kosterlitz Thouless transition point: model 
 (\ref{nien_triv}) does not allow one to enter the low temperature phase 
of 
 the XY model. Also, model (\ref{nien_triv}) has a first order transition
 for $N<-2$, which is not the behavior expected from the sigma model 
 analysis.

 It was suggested in \cite{JRS} that model (\ref{nien_triv}) can be 
 repaired by allowing for some intersections. The minimal scenario 
 one can imagine is to define a similar model on the square lattice, 
 and allow for self intersections at vertices only, so either none, 
 one or two loops go through the same vertex. The resulting objects are often
called trails. This gives the new 
 partition function
 \begin{equation*}
     Z_{T}=\sum_{{\cal G}'}K^{E}N^{L}w^{I}%\label{nien}
    \end{equation*}
where the sum is taken over all configurations $\mathcal{G}'$ of  closed loops,
which visit edges of the lattice at most once, in number $L$, 
covering a total of $E$ 
edges, with $I$ intersections.

The phase diagram of this model has not been entirely investigated.
It is expected that at least for $w$ small enough, the critical
behavior obtained with $K=K_{c},\,w=0$ is not changed (though $K_{c}$
is), while the low temperature behavior $K>K_{c}$ will be.

\begin{figure}
  \label{f:loop_cov}
\psfrag{b}{$B$}
\psfrag{c}{$C$}
\psfrag{e}{$E$}
\psfrag{o}{$O$}
  \centerline{\includegraphics{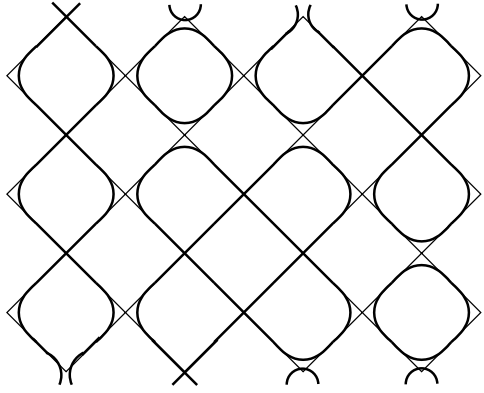}}
  \caption{Dense intersecting loop covering of a lattice with annulus
    boundary conditions. Illustration of bulk ($B$), contractible
    ($C$), even ($E$) and odd ($O$) loops. Periodic imaginary time runs vertically}
\end{figure}

In
\cite{JRS} a yet slightly different version was considered
corresponding, roughly, to the limit of very large $K$,
where all the edges of the lattice are covered. The partition function  of this fully packed  trails model then depends on only two parameters:
\begin{equation}
    Z_{FPT}=\sum_{{\cal G}''} N^{L}w^{I}\label{ours}
    \end{equation}
  and an example of allowed configuration is given in figure \ref{f:loop_cov}.   Numerical
and analytical arguments suggest strongly that this model has all the
generic properties of the $\OO(N)$ sigma  model in the spontaneously broken symmetry phase
(such that one might derive from analytic continuation in $N$ of the
RG equations),
for all $N\leq 2$.

Note that expression (\ref{ours}) can be obtained very naturally if instead of 
putting the degrees of freedom on the vertices, one puts them on the 
edges of the lattice. 
%\footnote{
%\textbf{I don't see how the intersecting dense loop model on a square 
%lattice can be reformulated with
%degrees of freedom on the sites a lattice. I mean, it is always 
%possible to switch the degrees of fredom to be on the sites of the 
%medial lattice, but they will no longer be spins taking discrete 
%values from 1 to $N$. }}
 In this case, the
minimal form of
interaction involves two edges crossing at one vertex. Invariance
under the $\OO(N)$ group allows for three invariant tensors as
illustrated on the figure \ref{f:vert}, while
isotropy  and
invariance under an overall scale change of the Boltzmann weights
leaves one with a single free parameter, the crossing weight $w$.
Graphical representation of the contractions on the invariant tensors
reproduces  eq.~(\ref{ours}), as will be discussed below.

\begin{figure}
  \psfrag{a}{a}
  \psfrag{b}{b}
  \psfrag{c}{c}
  \centerline{\includegraphics{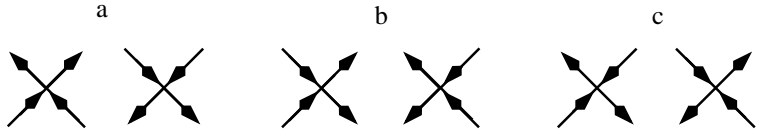}}
  \caption{Vertices of the $\mathbb{Z}_2$ symmetric six vertex
    model}\label{f:6v}
\end{figure}

\begin{figure}
  \psfrag{=}{$=$}
  \psfrag{+}{$+$}
 \centerline{\includegraphics{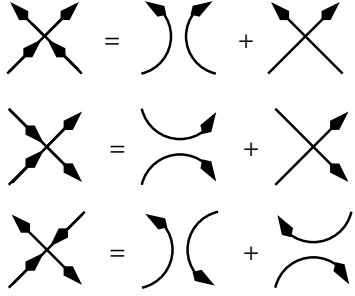}}
 \caption{The mapping of the six vertex model onto the oriented
loops}\label{f:6vO2map}
\end{figure}

For $N<2$, model (\ref{ours}) flows to weak coupling in the IR, and therefore
it is expected that the critical properties of the corresponding low temperature 
(Goldstone)
phase do not depend on  $w$, a fact checked numerically in
\cite{JRS}. The case $N=2$ is expected to be 
different:  as mentioned already in the introduction, the beta
function of the corresponding sigma model is exactly zero so the coupling
constant does not 
renormalize.
It is  indeed easy to see that the loop model  (\ref{ours}) with $N=2$ is equivalent to the  6 vertex model
with $a=b=1+w, \,c=1$.
Consider the vertices of the 6 vertex model as represented on figure
\ref{f:6v}. We chose isotropic weights $a=b,\, c$. We can decide to 
split the vertices of a configuration  into pieces of
oriented loops as represented on figure \ref{f:6vO2map}. For each
vertex, there are two possible splittings, and we assume that they
are chosen with equal probability. The loops
obtained by connecting all the pieces together provide a dense
covering of the lattice, and come with two possible orientations, 
hence a fugacity of two once the orientations are summed over. The 
loops can intersect, with a weight $w$ given from the obvious 
correspondence:
\begin{equation*}
    {a\over c}={b\over c}={1+w\over 2}
\end{equation*}
and thus
\begin{equation*}
    \Delta={a^{2}+b^{2}-c^{2}\over 2ab}=1-{2\over (1+w)^{2}}
\end{equation*}


We note that there are indeed {\sl three} invariant tensors for the
case of $\OO(2)$. The corresponding projectors are
$E,\, \tfrac{1}{2}(I-P),\, \tfrac{1}{2}(I-E+P)$. They
project respectively on two one dimensional representations, and on a single
two dimensional one.

The parameter  $\Delta$ covers the interval $ [-1,1]$ as $w\in [0,\infty]$. 
Changing $\Delta$ is well known to 
change  the exponents of  6 vertex  model, and therefore
eq.~(\ref{ours}) for $N=2$ exhibits a critical line, which  is in fact in the
universality class of the  continuous XY ($\OO(2)$) model in the low 
temperature (Kosterlitz Thouless) phase.

All what was said so far can be easily generalized to the case of 
spins taking values on a supersphere $\OSp(R|2S)/\OSp(R-1|2S)$. The 
fugacity of loops is now 
equal to $R-2S$: this combination is the number of bosonic minus the 
number of fermionic coordinates, and follows from the usual fact that 
when contracting fermions along a loop, a minus sign is 
generated~\footnote{The generalization of results for  $\OO(N)$ models 
to the case of orthosymplectic groups dates back to the work of 
Parisi 
and Sourlas \cite{ParSour}.}, see sec.~\ref{sec:1749}. The loop model formulation therefore 
provides a convenient graphical representation of the discrete 
supersphere 
sigma models for all $R-2S$, in particular $R-2S\leq 2$ where 
interesting physics is expected to occur. This physics was explored 
in \cite{JRS}, and the expected results were obtained for $R-2S<2$. The 
purpose of this paper is to explore the more challenging $R-2S=2$ 
case.

Of course, at the naive level of partition functions and without
worrying about boundary conditions, it looks as if there is no
difference between the
$\OO(N)$ spin model and its supersphere cousins provided $R-2S=N$.
The point is that the {\sl observables} of the
models are different or, at the very least, come with different
multiplicities. Indeed, consider for instance correlation
functions of spin variables.
In the $\OO(2)$ case, the spin has only
two components $S^{1},S^{2}$, so one cannot build a totally antisymmetric tensor on three indices. This means that the corresponding operator  
(which has a nice geometrical interpretation to be given in the next paper)
will not be present in this case, though it will be in the $\OSp(2S+2|2S)$ model when $S>0$. 
Note  that in general, correlators involving spins within 
the first $R$ bosonic and the first $2S$ fermionic labels 
 will be the same for {\sl any} choice of 
group $\OSp(R'|2S')$ with $R'-2S'=R-2S$ and $R'\geq R$.\footnote{Provided, of
 course, that the boundary conditions imposed on the $R'$ bosonic and $2S'$
fermionic degrees of freedom are the same in both cases.} This is 
immediately proved by performing a graphical expansion of the 
correlator: variables outside of the set of 
the first $R$ bosonic and the first $2S$ fermionic labels  are 
not getting contracted with the spins in the correlators, and cancel 
against each other in the loop contractions.

 A standard trick to extract the full operator content of a model is to
study the partition function with different boundary conditions.
Consider for instance the spin model on an annulus  with some symmetry preserving 
boundary conditions
in the space direction. With what we will call periodic boundary 
conditions (corresponding to taking the supertrace of the evolution 
operator)
%\footnote{\textbf{I do not agree with this statement.
%I remember we have agreed that antiperiodicity is defined by the 
%insertion of the matrix $J^{\otimes L}$} in the trace.
%Here $J$ is the invariant $\OSp{(R|2S)}$ form with the only nonzero 
%components, in the basis I prefer,
%$J_{i\bar{i}}=1$ and $J_{\bar{i}i} = (-1)^{\dg{(i)}}$.
%Imposing antiperiodicity will reduce to effectively setting $N=0$ in 
%the periodic (trace) partition function.
%If you want, one can give a boundary interpretation for the 
%supertrace by introducing
%an additional string perpendicular to all the others (at the top of a 
%Brauer diagram) with a fermion running on it.}
in the time
direction, representations of $\OSp(2S+2|2S)$ will always be counted with their superdimension, and the partition  
function will be identical with the one of the $\OO(2)$ case. 
But if we take antiperiodic boundary conditions, we will get a
modified partition function  (in the sense of \cite{ReadSaleur01})
which is a \emph{trace} over the Hilbert
space instead of a supertrace,  counts all observables with
the multiplicities (not supermultiplicities), and will turn out to be a very complex object.

A good algebraic understanding of the lattice model will be essential
 to make further progress, and, since the area is largely unexplored,
 this will occupy us for most of the rest of this first paper.

\section{Transfer matrices and algebra}

\subsection{Transfer matrices}\label{sec:spin}

As discussed briefly in the introduction, the  $\OSp(R|2S)$ spin model we consider is  most easily defined 
 on a square lattice with
degrees of freedom (states) on  the  edges and interactions taking 
place at  vertices.
The set of states on every edge is a copy of the base space
$V$ of the fundamental  $\OSp(R|2S)$ representation.
Interactions at a vertex can be encoded
in a local transfer matrix $t$
acting on $V^{\otimes 2}$ and commuting with the
$\OSp(R|2S)$ supergroup action.
We call $t$ an intertwiner and
write $t \in \End_{\OSp(R|2S)}V^{\otimes 2}$.

The Boltzmann weights
of the model are components of the transfer matrix
along a basis of intertwiners.
A natural choice of basis are the projectors onto $\OSp(R|2S)$
irreducible representations
appearing in the decomposition of the tensor product of two
fundamental  $\OSp(R|2S)$ representations.
To find them one can apply the same (anti)symmetrization and trace
substraction techniques used for reducing
$\OO(N)$ tensor representations. If $e_1,\dots,e_{R+2S}$ is a
mod 2 graded set of basis vectors in $V$ with grading $\dg$,
the decomposition of $V^{\otimes 2}$ will read
\begin{align}\label{eq:tens_prod}
  e_i\otimes e_j =&\frac{1}{2}\left(
    e_i\otimes e_j +    
    (-1)^{\dg(i)\dg(j)} e_j\otimes e_i -
    \frac{2J_{ij}}{R-2S} \sum_{k,l}J^{kl}e_k\otimes e_l
  \right) \\
  +& \frac{1}{2}\left(
    e_i\otimes e_j -
    (-1)^{\dg(i)\dg(j)} e_j\otimes e_i
  \right)
  + \frac{J_{ij}}{R-2S} \sum_{k,l}J^{kl}e_k\otimes e_l. \notag
\end{align}
Here $J_{ij}$ is the
$\OSp(R|2S)$ invariant tensor, $J^{ij}=(J^{-1})_{ij}$, and $g(i)=1$ 
(resp. $g(i)=0$)
if $i$ is fermionic (resp. bosonic).
Each of the three terms on the l.h.s.~of \eqref{eq:tens_prod}
transforms according to an irreps of $\OSp(R|2S)$, or,
in other words, belongs to a simple $\OSp(R|2S)$ module.

Introduce the identity $I$, the graded permutation operator $P$
(also known as braid operator),
and   $E$ the Temperley Lieb operator (proportional to the 
projector on the trivial representation),
\begin{equation}\label{e:osp_rep}
I^{\;\;kl}_{ij}=\delta^k_i \delta^l_j,\quad
P^{\;\;kl}_{ij} =(-1)^{\dg(i)\dg(j)}\delta^k_i\delta^l_j, \quad
E^{\;\;kl}_{ij} =J^{kl}J_{ij}.
\end{equation}
In terms of projectors onto irreducible $\OSp(R|2S)$ modules,
eq.~\eqref{eq:tens_prod} may be written in a more elegant way as
\begin{equation*}\label{e:tp}
I=\frac{1}{2} \left( I + P - \frac{2}{R-2S} E \right) +
\frac{1}{2}\,(I-P) + \frac{1}{R-2S}\,E.
\end{equation*}

Let $\mathsf{P}$ denote as usual the inversion of space, $\mathsf{T}$
the inversion of time and $\mathsf{C}$ the charge conjugation with the
matrix $J$.
One can check directly from definition \eqref{e:osp_rep} that $P$ is
$\mathsf{C}_{12}$,  $\mathsf{P}$ and
$\mathsf{T}_{12}$ invariant, while
$E$ is  $\mathsf{P}_{12}$ and $\mathsf{C}_{12}\mathsf{T}_{12}$
invariant.
Moreover, $E$ and $I$ transform into each other under the $\pi/2$
rotation of the lattice $\mathsf{R}$,
while $E$ and $P$ are related by the crossing symmetry
$\mathsf{C}_1\mathsf{T}_1$
\begin{align*}
  \mathsf{R}&: \; E^{\;\;kl}_{ij}\; \longrightarrow \;
  J_{jj'}E^{\;\;lj'}_{k'i}J^{k'k}
  =I^{\;\;kl}_{ij} \\
  \mathsf{C}_1\mathsf{T}_1&: \; E^{\;\;kl}_{ij}\; \longrightarrow \;
  J_{ii'}E^{\;\;i'l}_{k'j}J^{k'k}
  =P^{\;\;kl}_{ij}.  
\end{align*} 

Take $I$, $E$ and $P$ as basis of intertwiners in
$\End_{\OSp(R|2S)}V^{\otimes 2}$.
The local transfer matrix generally depends on three independent
weights $w_I$, $w_E$ and $w_P$. However,
on a homogeneous and isotropic lattice
one can normalize $w_I=w_E=1$ and leave only the weight $w= w_P$.
Finally, the local transfer matrix takes the form
\begin{equation}\label{e:loc_trmat}
t(w)=I + w P + E.
\end{equation}
\begin{figure}
  \psfrag{x}{$Y$}
  \psfrag{y}{$X$}
  \centerline{\includegraphics{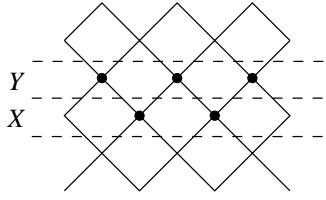}}
  \caption{The one layer transfer matrices
    $X$ and $Y$ represented on a diagonal lattice of width 6.}
  \label{f:dl}
\end{figure}

On a diagonal lattice with open boundaries represented in
fig.~\ref{f:dl} choose the time in vertical direction.
The notation of sites at a fixed time is
such that the left edge $i$ and right edge $i+1$ meet at vertex $i$.
Let $t_i(w)\in \End_{\OSp(R|2S)}V^{\otimes L}$ denote
a transfer matrix acting nontrivially only at vertex $i$
according to eq.~\eqref{e:loc_trmat}.
From the figure it is clear that odd and even
times are inequivalent.
The transfer matrix $T$, propagating one step forward at equivalent
times, may be written as a product $T=YX$ of one layer transfer
matrices
\begin{equation}\label{e:xy}
X = \prod_{i=1}^{[(L-1)/2]}t_{2i},\quad Y = \prod_{i=1}^{[L/2]}t_{2i-1},
\end{equation}
schematically shown in fig.~\ref{f:dl}.

The simplest way to define a partition function that depends
on the whole spectrum of
the transfer matrix $T$
is by taking the \emph{trace} of $T$
at a certain power $\beta$.
Selecting other boundary conditions with some nontrivial symmetry
generally amounts to 
restricting the whole space of states of the model to a subspace 
compatible with the symmetry of chosen boundary
conditions.
What exactly we mean by ``symmetry of boundary
conditions'' will be explained later in sec.~\ref{sec:1749}.
For the moment let us just say that
it is convenient to consider a more general class of boundary conditions,
called quasiperiodic, in which $T^\beta$ is
``twisted'' by the action of an element $D$
of the supergroup.
Define the quasiperiodic partition function to be
\begin{equation}
  \label{eq:qppf}
  Z_D = \str_{V^{\otimes L}} D^{\otimes L} T^\beta. 
\end{equation}
We must take the \emph{supertrace} in eq.~\eqref{eq:qppf}
if we want the quasiperiodic partition function to be well defined.
For instance, when $D=J^2$ we get the usual trace partition function and when $D$
equals to the identity matrix we get the supertrace partition function. \\
Note that because $D$ is a
supermatrix, the tensor product in  $D^{\otimes L}$ has to be graded,
that is
\begin{equation*}
  \label{eq:super_tp}
  D^{\otimes 2} \cdot \eta\otimes \xi = D\cdot \eta \otimes D\cdot \xi
  \quad \Rightarrow \quad 
  \left(D^{\otimes 2}\right)_{\;\; kl}^{ij}
  =(-1)^{\dg(k)(\dg(j)+\dg(l))} D^i_{\;k} D^j_{\;l}
\end{equation*}

After inserting the local transfer matrix from eq.~\eqref{e:loc_trmat}
in eq.~\eqref{e:xy} and expanding the transfer matrix $T$, the 
quasiperiodic partition function 
reads as a sum of weighted
products of $E_i$'s and $P_i$'s.
Such linearly independent products must be considered as
words of a \emph{transfer matrix algebra}, while intertwiners $E_i$
and $P_i$ are \emph{generators} of this algebra.
In the next section we identify this algebra as a 
representation of the Brauer algebra.

\subsection{The Brauer algebra} \label{sec:br_alg}

For an abstract introduction to the Brauer algebra see ref.~\cite{Ram95,
  Hanlon99} while in the context of $\osp(R|2S)$ centralizer algebra see
ref.~\cite{Ram98}. We collect in this section some well known facts
about the Brauer algebra we shall use in the next sections.

Let $E_i$ and $P_i$, $i=1,\dots, L$ act nontrivially as $E$ and $P$
in eq.~\eqref{e:osp_rep} only at the sites
$V_i\otimes V_{i+1}$ of $V^{\otimes L}$.
One can check that for $P_i$
and $E_i$ so defined the following relations hold:
\begin{align}\label{e:def_rel}
P_i^2 =1,\quad E_i^2=N E_i, \quad E_i P_i = P_i E_i = E_i,\\ \notag
P_i P_j = P_jP_i, \quad E_i E_j = E_j E_i, \quad E_iP_j = E_j P_i,\\ 
\notag
P_iP_{i \pm 1}P_i=P_{i\pm 1}P_i P_{i \pm 1}, \quad E_i E_{i \pm 1} 
E_i = E_i, \\ \notag
P_i E_{i\pm 1}E_i = P_{i \pm 1} E_i, \quad E_{i} E_{i\pm 1} P_i = E_i
P_{i \pm 1}. 
\end{align}
In the second line of these relations $i$ and $j$ are 
supposed to be nonadjacent sites.

Relations (\ref{e:def_rel}) (is one of the many ways to) define the  $B_L(N)$
Brauer algebra (also denoted sometimes by 
the names of braid-monoid algebra or degenerate 
Birman-Wenzel-Murakami algebra \cite{BW,Mu,WDA}).
Note that this algebra depends
on a single, generally complex, parameter $N$, and 
contains the maybe more familiar Temperley Lieb 
algebra, generated by $E_i$'s alone, and the symmetric
group algebra, generated by $P_i$ alone.

For  $N$ fixed and $L$ big enough, the $\OSp(R|2S)$ spin models
provide highly unfaithful representations of the Brauer algebra $B_L(N)$.
This is because, in  $V^{\otimes L}$, 
the generators $P_i$ and $E_i$  
satisfy additional 
higher order relations $\mathcal{R}$ on top of \eqref{e:def_rel}.\footnote{This 
situation is similar to what happens for models with $\SL(N)$ 
symmetry in the fundamental representation, and corresponding quotients 
of the Hecke algebra.} For a simple example, consider  the $\OO(2)$
spin model on a lattice of width 3. The projector onto the
antisymmetric tensor of rank 3 is zero, thus, $\mathcal{R}$ contains
the additional relation $1+P_1P_2+P_2P_1 = P_1+P_2+P_1P_2P_1$.
Our spin models in general provide  representations of the
quotient algebras $B_L(N)/\mathcal{R}$.
The set of relations $\mathcal{R}$ can be explicitly described for
$S=0$, see \cite{Weyl_book,Gavarini06} and references therein, and we 
have little to say about the case $S>0$.

The first step in understanding the spectrum of the transfer matrix
$T$ brings up the question of $B_L(N)$ irreducible representations,
and  of their multiplicities in $T$
for a particular choice of $R$ and $S$.
This leads us to discussing some results about the representation 
theory of the Brauer algebra.

\begin{figure}
 \centerline{\includegraphics{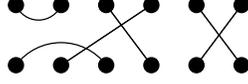}}
 \caption{The graphical representation of the word $P_5P_3E_1P_2$ in
   $B_6(N)$.}
 \label{f:1factor}
\end{figure}

The most natural representation to begin with  is the {\sl adjoint
  representation}.
It admits a diagrammatic representation in terms of
graphs on $2L$ points in which every vertex has degree 1.
Usually one orders the $2L$ points on two horizontal 
parallel lines as shown in fig.~\ref{f:1factor}.
Let $B_L$ denote the vector space spanned on the $(2L-1)!!$ such
diagrams.

\begin{figure}
  \psfrag{1}{1}
  \psfrag{L}{$L$}
  \psfrag{i}{$i$}
  \psfrag{j}{$i+1$}
  \psfrag{I}{$I$}
  \psfrag{E}{$E_i$}
  \psfrag{P}{$P_i$}
 \centerline{\includegraphics{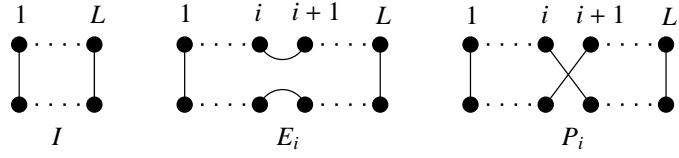}}
 \caption{Graphical representation of generators $I$, $E_i$ and $P_i$.}
 \label{f:gen_diag}
\end{figure}

The product $d_1*d_2$ of two diagrams $d_1$ and $d_2$ is performed by
putting $d_1$ on top of $d_2$ and replacing each of the loops
in the resulting diagram with $N$.
Define diagrammatically the identity $I$  and the generators $E_i$, $P_i$
as represented in fig.~\ref{f:gen_diag}. One can check that the
graphical representation of generators with the multiplication $*$
of diagrams satisfy all of the eqs.~\eqref{e:def_rel}. 
The left action of
generators on
$B_L$ via the multiplication $*$ of diagrams provide the adjoint
representation of $B_L(N)$.

From the graphical representation we see that 
$B_L$ has a series of invariant subspaces
$B_L=B^L_L \supset B^{L-2}_L\supset \dots \supset B^\tau_L$,
where $B^m_L$ is spanned on diagrams with fewer than $m$ vertical
lines and $\tau=L\mod 2$.
The vector space spanned on diagrams with
exactly $m$ vertical lines may be defined as a $B_L(N)$ module by the coset
$B'^m_L =B^m_L/B^{m-2}_L$.
The left action of $B_L(N)$ on this modules may be seen as a
modified multiplication $*_m$ of diagrams
\begin{equation*}
  \label{eq:mod_prod}
  d_1 *_m d_2 = \begin{cases}
    d_1 * d_2, &\text{if it has $m$ vertical lines} \\
    0,&\text{otherwise}
    \end{cases}
\end{equation*}

\begin{figure}
  \psfrag{1}{1}
  \psfrag{2}{2}
  \psfrag{3}{3}
  \psfrag{4}{4}
  \centerline{\includegraphics{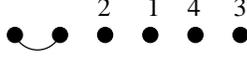}}
 \caption{The top of the diagram represented in fig.~\ref{f:1factor}.
   The labeling permutation is $(2,1,4,3)$.}
 \label{f:lab_1fact}
\end{figure}

Under the left action of the algebra the position of horizontal lines
in the bottom of a diagram does not change.
For a given configuration of the
horizontal lines in the top of a diagram and a given 
pattern of intersections of vertical lines
there are $(L-m-1)!!C_L^m $ possibilities of
choosing the configuration of horizontal lines in the bottom of the
diagram.
This simply means that  $B'^m_L$ decomposes into a direct sum of $(L-m-1)!!C^m_L $ equivalent
modules.
The coset representative $B''^m_L$ of these equivalent left
modules is spanned
on $m!(L-m-1)!!C^m_L$ graphs on $L$ points with
every vertex having degree 0 or 1 and a labeling
with numbers $1,\dots,m$ of free vertices.
An example of such a labeled graph is shown in
fig.~\ref{f:lab_1fact}.
If the labellings are omitted the resulting graph is called a partial
diagram.

The labeling of the $m$ free points of a labeled graph is a
permutation $\pi$ in the symmetric group $\Sym(m)$.
The labeled graphs will provide a representation of the
Brauer algebra, which is irreducible for generic values of $N$, if we take
the labellings $\pi$ in an irreducible
representation of $\Sym(m)$.
We call such representations generically irreducible.
Let $\mu$ be a partition of $m$, which we write as $\mu\vdash m$.
In a more algebraic language the definition of generically irreducible
left modules translates to
\begin{equation}\label{e:gen_irr}
\Delta_L(\mu):= B''^m_L \otimes_{\Sym(m)} S(\mu),~~~\mu\vdash m,
\end{equation}
where $S(\mu)$ is an irreducible $\Sym(m)$ module.
In view of later numerical analysis we give below a basis
in $\Delta_L(\mu)$ and describe the action of $B_L(N)$ on this basis.

Let $p\otimes \pi$ denote the labeling of a partial diagram $p$ with
the permutation $\pi$,  $v_1,\dots,v_{f_\mu}$ be a set
of basis vectors in $S(\mu)$ and $\rho_\mu(\sigma)$ be the matrix
of the permutation $\sigma$ in the representation $\rho_\mu$.
A natural basis in $\Delta_L(\mu)$ is given by all pairs
$p\otimes v_i$.
The action of a diagram $d\in B_L(N)$ on a basis vector is
\begin{equation}
  \label{eq:action}
  d \cdot p\otimes v_i=\begin{cases}
      d*_m p \otimes  \rho_\mu(\sigma^{-1})v_i,
      &\text{if $d*p$ has $m$
	free points} \\
      0, &\text{otherwise},
  \end{cases}
\end{equation}
where $\sigma$ is the labeling of $d*g$ and
$g$ is the partial diagram $p$ labeled with the identity permutation.
The dimensions $d_\mu$ of $\Delta_L(\mu)$ is
$f_\mu (L-m-1)!! C_L^m$.

In simple words,  a generically irreducible module is a span on graphs
on $L$ points,
obtained by choosing $m$
points among $L$, pairing all the others (this
gives the multiplicity $(L-m-1)!!$ since intersections are allowed),
choosing for the $m$ unpaired ones a representation of the permutation
group and setting to zero the action of any Brauer diagram
that reduces the number $m$ of unpaired points.

The generically irreducible representations labeled by $\mu\vdash
L-2k$, $k=0,\dots,[L/2]$
appear in the decomposition of the adjoint representation
with  multiplicity given by their dimension $d_\mu$ when $B_L(N)$ is semisimple.

Let us conclude with a few words about the reducibility of generically
irreducible modules $\Delta_L(\mu)$.
For integer $N$ and a number of
strings $L>N$ the Brauer algebra is not semisimple and, as a
consequence, certain of the modules $\Delta_L(\mu)$ become reducible,
though they remain \emph{indecomposable}.\footnote{We adopt here
the physicist's habit of calling indecomposable a module which
is reducible though not fully reducible. Therefore the set
of indecomposables does not contain the irreducibles,
unlike in most of the math literature.}
The irreducible components 
appearing in such reducible  modules $\Delta_L(\mu)$ are
far from being understood (the situation 
is much worse than in the case of the  nonsemisimple Temperley
Lieb algebra \cite{Jones,Martin}).
Numerical computations based on the diagonalization of the
transfer matrix in the diagrammatic representation of $B_L(N)$
restricted to $\Delta_L(\mu)$ decreases
in efficiency very fast with increasing $L$, compared to the ideal case where
the transfer matrix is restricted to an irreps of $B_L(N)$. 
This is because
for big $L$ and $\mu$ fixed the number of irreducible components in
$\Delta_L(\mu)$ ``goes wild'' and there are a lot of ``accidental degeneracies''
in the spectrum of the transfer matrix restricted to $\Delta_L(\mu)$.

However, a significant progress in this direction has been recently
made in \cite{Martin06,Martin12/06}. Let us note that, as described in
\cite{Martin06},  the content of (at least some) $\Delta_L(\mu)$'s can
be computed by
repeated applications of Frobenius reciprocity applied to the short
exact sequence of \cite{Hanlon99}
describing the structure of the induced modules
$B_{L+1}(N)\otimes_{B_L(N)}\Delta_L(\mu)$.

In the end we recall the basic results for the Temperley Lieb algebra,
to allow a quick comparison with Brauer.
Temperley Lieb algebra diagrams are a subset of Brauer algebra
diagrams subject to the constraint that no intersections between edges
are allowed.
The dimension of the algebra is given by
the Catalan numbers $(2L)!/L!(L+1)!$.
The main line of reasoning for
finding generically irreducible modules follows the same way,
except there is no available action of the symmetric group
on vertical lines.
Therefore, the analogue of the labeled graphs will be the partial
diagrams, in which no free points may be trapped
inside an edge.
The number of such graphs is $C^n_{L-1} -
C^{n-2}_{L-1}$, where $n=(L-m)/2$ is the number of edges.
The generically irreducible modules $D_L(m)$ are parametrized by the number $m
= L, L-2,\dots$ of free points in the graphs.

The presented facts about the Brauer algebra should be enough
to understand the loop gas reformulation of $\OSp(R|2S)$ spin model,
which we give in the next section.

\subsection{Loop reformulation of $\OSp(R|2S)$ spin  models: the 
algebraic point of view}
\label{sec:1749}

\begin{figure}
  \psfrag{=}{$=$}
  \psfrag{+}{$+$}
  \psfrag{I}{$I$}
  \psfrag{E}{$E$}
  \psfrag{P}{$P$}
  \psfrag{t}{$t$}
  \psfrag{1}{1}
  \psfrag{w}{$w$}
  \centering
  \includegraphics{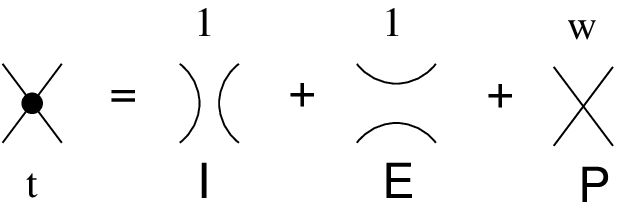}
  \caption{Possible vertex of interactions in the loop model.}
\label{f:vert}
\end{figure}

The emergence  of dense intersecting loops
becomes transparent if we take the local transfer matrices in
the adjoint representation of the Brauer algebra.
This simply amounts to replacing in eq.~\eqref{e:loc_trmat}
the generators $I$, $E_i$ and $P_i$ defined by eq.~\eqref{e:osp_rep}
with the diagrams in fig.~\ref{f:gen_diag}.
The adjoint local transfer matrix is represented in
fig.~\ref{f:vert}.

\begin{figure}
  \psfrag{=}{$=$}
  \psfrag{N}{$N$}
  \centering
  \includegraphics{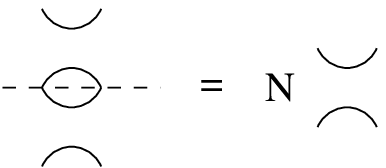}
  \caption{Emergence of bulk loops.}
\label{f:e2}
\end{figure}

We now define a loop model on a diagonal lattice represented in
fig.~\ref{f:dl}, 
with reflecting boundaries on the left and right (ie, free boundary 
conditions in the space direction)
and identified boundaries on the top and bottom (quasiperiodic boundary 
conditions in the time direction.)
The states of
this model are coverings of the lattice with dense
\emph{intersecting} loops.
Dense means that every edge on the lattice
necessarily belongs to a loop.
Avoiding loop vertices have weight 1 and  intersections come with 
weight $w$.
There are two possible ways for a line to close in a
loop.
The first one comes from the graphical representation of the
relation $E^2=N E$ in fig.~\ref{f:e2}.
We call such loops \emph{bulk} loops.
Clearly the fugacity of bulk loops is fixed to $N$ by the Brauer algebra.
The second possibility is that the ends of the line close in the identified
points of the top and bottom boundaries of the lattice.
We call such loops \emph{cycles}.
The boundary condition we consider have an annulus geometry and, thus,
a cycle can be either \emph{contractible} or \emph{uncontractible}.
The fugacity of cycles is not fixed by the algebra.
In fact, as we explain below,
this is exactly the degree of freedom allowing for
multiple mappings from the $\OSp(R|2S)$ spin models with $R-2S=N$ fixed and
the dense intersecting loop model with fugacity
$N$ for loops.

We start by evaluating the trace $\tr_{V^{\otimes L}}d$ of a
diagram $d$ in the spin representation and then we generalize the
result for quasiperiodic boundary conditions given by
$\str_{V^{\otimes L}}D^{\otimes L} d$. We follow the same line of reasoning as
in \cite{Ram95}.

A  cycle in a diagram $d$ is the subgraph on the
set of points belonging to a loop if
we identify its top and bottom vertices.
By an abuse of language we call the corresponding loop also cycle.
If we put a diagram $d_1$ to the left of a diagram $d_2$ we get a new
diagram which we denote $d_1\otimes d_2$.
Let $c_1,\dots,c_l$ be the cycles in $d$.
We can separate them by permuting the
top and bottom vertices of $d$ with the same permutation $\pi$
\begin{equation*} 
 \pi* d *\pi^{-1}= c_1\otimes\dots\otimes c_l.
\end{equation*}
Thus the trace of a diagram depends only on the weights of cycles
\begin{equation*}
  \tr_{V^{\otimes L}}d = \tr c_1 \dots \tr c_l.
\end{equation*}
More than that, the weight of a cycle
depends only on how many times it winds the annulus.

Indeed, if a cycle on $2m$ points has no horizontal lines,
then, by applying the same permutation to the top and bottom vertices
we can bring it to the cycle $P_1\dots P_{m-1}$.
This is because permutations with one cycle are conjugate in
$\Sym(m)$.

If a cycle $c$ has a horizontal edge between the first
and the second vertex in the top then it has the same weight as a
certain cycle $c'$ on four points less then $c$
\begin{equation}\label{eq:contr}
  \tr c = \frac{1}{N}\tr E_1 * c = \frac{1}{N} \tr c * E_1 =
  \frac{1}{N} \tr E_1 \otimes c' =\tr c'.
\end{equation}
If we compare the $c$ on the left with $c'$ on the right 
\begin{figure}[h]
  \centering
  \includegraphics{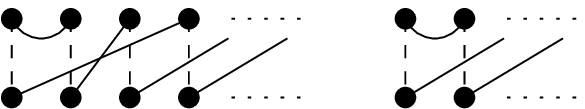}
\end{figure}
it is clear that, in the end of the iterative application of
eq.~\eqref{eq:contr}, the final cycle can be interpreted as
being the initial cycle $c$ maximally contracted on the annulus.

In the end, the only weights we need to compute explicitly are 
that of the cycles $E$ and $P_1\dots P_{m-1}$
\begin{align*}%\label{eq:tre}
  &\tr_{V^{\otimes 2}} E =  J_{i_1i_2}J^{i_1i_2}= N \\ %\label{eq:trp}
  &\tr_{V^{\otimes m}}P_1 \dots P_{m-1} =
  (-1)^{\dg(i_1)(\dg(i_2)+\dots+\dg(i_m))}\delta_{i_1=\dots=i_m} =
  R+(-1)^{m+1}2S.
\end{align*}

For boundary conditions twisted by the matrix $D\in \OSp(R|2S)$ 
the generalized weights are computed to be 
\begin{align}\notag
 &\str_{V^{\otimes 2}} D^{\otimes 2}E = N \\ \label{eq:cyc_w}
 &\str_{V^{\otimes m}}D^{\otimes m}P_1 \dots P_{m-1} =
 \str D^m.
\end{align}

In the fundamental representation, every supermatrix $D$ is
diagonalizable.
The diagonal form of $D\in \OSp^+(R|2S)$ in the
fundamental representation is determined
by exponentiating elementary weights $\epsilon_i$ and $\delta_j$
introduced in sec.~\ref{sec:superalg}.
Thus, $D$ restricted to $V_0$ has eigenvalues
$x_1,x^{-1}_1,\dots,x_r,x^{-1}_r,(x_{r+1}=1)$ 
and restricted to $V_1$ has eigenvalues $y_1,y^{-1}_1,\dots,y_S,y^{-1}_S$.
The braces in $(x_{r+1})$ mean that $x_{r+1}$ appears for odd $R$ only.
Eq.~\eqref{eq:cyc_w} can now be rewritten
\begin{equation}
  \label{eq:trp+}
  \str D^m = \sum_{i=1}^r \big( x_i^m + x_i^{-m} \big)+ (1) - \sum_{j=1}^{S}
  \big( y_j^m+y_j^{-m}\big).
\end{equation}
For $D\in \OSp^-(R|2S)$ only the eigenvalues in $V_0$ change with
respect to the previous case.
There are of the form
$x_1,x_1^{-1},x_2,x^{-1}_2,\dots,x_r,x^{-1}_r,x_{r+1}=-1$
for $R$ odd, while for $R$ even
$x_1,x_1^{-1},x_2,x^{-1}_2,\dots,x_{r-1},x^{-1}_{r-1}$ and
$x_r=1,x_r'=-1$.
Instead of eq.~\eqref{eq:trp+} one has now
\begin{align*}
  %\label{eq:trp-}
   \str D^m &= \sum_{i=1}^r \big(x_i^m + x_i^{-m}\big) +(-1)^m - \sum_{j=1}^{S}
  \big(y_j^m+y_j^{-m}\big), \quad& R\textnormal{ odd}\\
   \str D^m &= \sum_{i=1}^{r-1} \big(x_i^m + x_i^{-m}\big) +1+(-1)^m - \sum_{j=1}^{S}
  \big(y_j^m+y_j^{-m}\big), \quad& R\textnormal{ even}.
\end{align*}

To summarize the basic results in this section,
let $\mathcal{G}$ be a dense loop covering of the lattice,
$I$ be the number of intersections, $B$ be the
number of bulk loops, $C$ be the number of contractible loops (cycles) and
$E$($O$) be the number of loops winding the annulus an
even(odd) number
of times.

On the annulus the trace partition function (which would correspond to
antiperiodic boundary conditions in the (imaginary) time direction) of the
$\OSp(R|2S)$ spin
model may be reformulated as a dense intersecting loop model in
the following way
\begin{equation}
  \label{eq:loop_pf}
  Z= \sum_{\mathcal{G}} w^I N^{B+C + E}(R+2S)^O.
\end{equation}
We see that it does depend on $R,S$ separately  and not only on $N$.

Meanwhile the supertrace partition function  (which would correspond to periodic couplings) reads
\begin{equation*}
  Z= \sum_{\mathcal{G}} w^I N^L,
\end{equation*}
where $L=B+C+E+O$ is the total number of ``loops'' and, since it
depends on $N$ only, is the same as for the $\OO(N)$ model. One can say that
taking the supertrace in the partition function
is equivalent to restricting the $\OSp(R|2S)$ supersymmetry of the
spin model to a, smaller, $\OO(N)$ symmetry.

Denote the spectrum of the transfer
matrix of the $\OSp(R|2S)$ spin model by $\Sigma_S(N)$.
We have the following important inclusion property
\begin{equation}
  \label{eq:filtr}
  \Sigma_0(N) \subset \Sigma_1(N) \subset \Sigma_2(N) \subset \dots.
\end{equation}
The only difference between the $\OSp(R|2S)$
and $\OSp(R-2|2S-2)$ quasiperiodic partition functions is the weight of
uncontractible cycles.
For $D\in \OSp(R|2S)$ a matrix with eigenvalues
$x_i = 1$ and $y_j=-1$, except $y_1=1$, the weight of
uncontractible cycles is, according to
eq.~\eqref{eq:trp+} either  $N$ or $R+2S-4$.
Notice that these are exactly the weights of
uncontractible cycles in the trace partition function for the
$\OSp(R-2|2S-2)$ spin model, which proves eq.~\eqref{eq:filtr}.

% It is easy to see from eq.~\eqref{eq:trp+} that the supertrace
% partition function of the $\OSp(R|2S)$ spin model is the same as the trace
% partition function of the $\OO(N)$ spin model.

% 
% One the other hand, taking $D=\rho$ has the effect of
% giving a fugacity $N$ to all loops except those winding the cylinder an
% odd number of
% times. The latter get a fugacity $N-2$.

We will use the inclusion property \eqref{eq:filtr} in
the next section to derive some information about the
indecomposable representations of
$\OSp(R|2S)$ appearing in the decomposition of $V^{\otimes L}$.

\section{Decomposition of $V^{\otimes L}$}

\subsection{ General results}
\label{sec:gf}

Assume the transfer matrix be a generic element  
of $B_L(N)$.
The action of $B_L(N)$ on the tensor space $V^{\otimes L}$ was defined
in the beginning of sec.~\ref{sec:br_alg}.
The complete picture of the reducibility of the transfer matrix can
be conveniently encoded in the decomposition of $V^{\otimes
  L}$ into a direct sum of $B_L(N)$ indecomposable modules
\begin{equation}
\label{eq:bd}
  V^{\otimes L}\simeq \bigoplus_{\lambda \in Y_L(S)} m_\lambda
  \mathcal{I}B_L(\lambda),
\end{equation}
where $m_\lambda$ denotes the multiplicity of
isomorphic indecomposable
$B_L(N)$ modules $\mathcal{I}B_L(\lambda)$ (it does not depend on $L$), and the set $Y_L(S)$ 
is defined implicitly by the formula, and will be defined 
explicitly below.
We remind the reader that $V^{\otimes L}$ is not necessarily
a semisimple $B_L(N)$ module if $L>N$, so the modules
$\mathcal{I}B_L(\lambda)$ appearing on the rhs of eq.~\eqref{eq:bd}
can be reducible. 

The question of computing degeneracies of eigenvalues of the spin transfer
matrix is easier to treat by looking at the centralizer
$\mathsf{Z}:= \End_{B_L(N)} V^{\otimes L}$, which acts on
$V^{\otimes L}$ from the left if one consider $B_L(N)$ acting
from the right.
The dimension of indecomposable modules $\mathcal{I}G(\mu)$
in the decomposition of $V^{\otimes L}$ as a $\mathsf{Z}$-module
\begin{equation}
  \label{eq:cd}
    V^{\otimes L}\simeq \bigoplus_{\mu \in X_L(S)} n^\mu_L \mathcal{I}G(\mu)
\end{equation}
will give the desired degeneracies.
This is due to the fact
that $n^\mu_L$ are dimensions of
simple $B_L(N)$ modules $B_L(\mu)$ appearing as constituents of
$\mathcal{I}B_L(\lambda)$ in  eq.~\eqref{eq:bd}, while $m_\lambda$ are
dimensions of simple $G$ modules $G(\lambda)$ appearing as constituents of
$\mathcal{I}G(\mu)$ in  eq.~\eqref{eq:cd}.
Taking the character of both eqs.~(\ref{eq:bd},\ref{eq:cd}) one can
see that the number $b(\lambda,\mu)$ of irreducible components
$B_L(\mu)$ in $\mathcal{I}B_L(\lambda)$ is equal to the number
$g(\mu,\lambda)$ of irreducible components
$G(\lambda)$ in $\mathcal{I}G(\mu)$.

Because the action of $\osp(R|2S)$ commutes with $B_L(N)$
we have that $\osp(R|2S)\subset \mathsf{Z}$.
However, when  $V^{\otimes L}$ is semisimple it follows from the Wedderburn
decomposition theorem that  $\mathsf{Z} \simeq
\mathbb{Z}_2\times \osp(R|2S)$.
In the following we suppose that there is still a Schur duality between
 $\osp(R|2S)$  and the quotient of $B_L(N)$ faithfully represented on
 $V^{\otimes L}$.
This allows us to give an algorithm to compute the lhs of
eqs.~(\ref{eq:bd},\ref{eq:cd}) for small tensor powers $L$ and get
some intuition about
the general structure of $\mathcal{I}B_L(\lambda)$ and
$\mathcal{I}G(\mu)$.

The set of partitions $X_L=\{\,\mu\vdash L-2k\mid k=0,\dots,[L/2]\,\}$
labels $B_L(N)$ irreps, while
$X_L(S)\subset X_L$ selects those of them
which do realize on the tensor space
$V^{\otimes L}$.
Denote by $J(S)\subset B_L(N)$ the double sided ideal defined by
$V^{\otimes L}\cdot J(S)=0$.
The annihilator $J(0)$ is diagrammatically described in 
\cite{Gavarini06}.
Under the homomorphism $\rho:B_L(N)\rightarrow B_L(N)\,/J(S)$, the
indecomposable modules $\Delta_L(\mu)$ give rise to induced 
modules $\delta_L(\mu)=\Delta_L(\mu)\,/J(S)\cdot\Delta_L(\mu)$.
Clearly, $\delta_L(\mu)$ is a tensor representation and can be
generated by trace substraction and symmetrization as
\begin{equation}
\label{eq:gend}
  V^{\otimes L} \mathcal{T}_{L-2k} e_\mu E_{L-2k} \dots E_{L-1},
\end{equation}
where $\mu\vdash L-2k$, $\mathcal{T}_{L-2k}\in B_L(N)$ extracts all
the traces from the tensor space $V^{\otimes L-2k}$ 
and $e_\mu$ acts nontrivially only on $V^{\otimes L-2k}$ as a Young
symmetrizer.
%Explicit examples of such tensor modules are given in the next section.
The double sided ideal $J(S)$ is completely characterized by the set of
weights $ X_L(S)=\{\, \mu \in X_L\mid J(S) \cdot B_L(\mu)=0\,\}$.
Note that $X_L(S)\subset X_{L+2k}(S), k\geq 1$.
The surviving indecomposable tensor modules $\delta_L(\mu)$
are given by  $\Delta_L(\mu)$ with irreducible components $B_L(\nu),\nu\notin X_L(S)$
removed.
The quotient $B_L(N)/J(S)$ can be carried out by imposing the vanishing
of all words $W_\mu:= \mathcal{T}_{L-2k} e_\mu \in B_L(N)$ with $\mu \in
X_L/X_L(S)$. 
It is useful to notice that not all of these conditions are
independent and as one can see from eq.~\eqref{eq:gend} $W_\mu =0
\Rightarrow W_\nu = 0$ if $\mu\subset \lambda$.

As discussed in sec.~\ref{sec:superalg} and \ref{sec:supergr},
the $\osp(R|2S)$ irreducible components of $\mathcal{I}G(\mu)$ are
indexed  (up to an equivalence under the action of the outer
automorphism $\tau$ induced by the symmetry of the Dynkin diagram of
$\osp(R|2S)$ when $R$ even) by the set
$H_L(S)=\{\,\lambda\in X_L(S)\mid \lambda_{r+1}\leq S\,\}$ of hook
shape partitions.
Representing the supergroup as a semidirect product
$\OSp(R|2S)=\mathbb{Z}_2 \times
\OSp^+(R|2S)$, the elements of $Y_L(S)$ naturally acquire the
structure of a couple of the form $1\times \lambda$ or 
$\varepsilon\times \lambda$ if
$\lambda_S<r$ and $\tau\times
\lambda$ if $\lambda_S\geq r$,
where $1,\varepsilon,\tau$ are the
trivial, alternating (superdeterminant)  and  two
dimensional representations of 
$\mathbb{Z}_2=\OSp(R|2S)/\OSp^+(R|2S)$.
Thus, every $\lambda\in H_L(S)$ gives rise
to two $\OSp(R|2S)$ inequivalent irreps with highest weights
$\lambda:= 1\times \lambda$ and the associate
$\lambda^*:=\varepsilon \times\lambda$ if $\lambda_S<r$ and
a single self-associate irreps of highest weight
$\lambda=\lambda^*:= \tau\times
\lambda$ if $\lambda_S\geq r$.
For typical $\lambda\in H_L(S)$, one can realize the
$\OSp(R|2S)$ irreps $\lambda,\lambda^*$ on tensors $T_L(\lambda),T_L(\lambda^*)$
and describe their symmetry by some Young tableaux.
As discussed in details in sec.~\ref{sec:tmb}, the Young tableau
corresponding to $T_L(\lambda^*)$ can be constructed by adding a border
strip to the Young tableau of shape $\lambda$ corresponding to
$T_L(\lambda)$.
Ultimately, this is justified by the fact that
eq.~(\ref{eq:elchar},\ref{eq:onchar}) gives the right 
characters for $T_L(\lambda), T_L(\lambda^*)$ and that they coincide up to $\sdet D$. 
Although atypical representations cannot be realized as tensor
representations we
represent the associate weight $\lambda^*$  of an  atypical weight
$\lambda$ by a Young tableau such that $\lambda^*/\lambda$ is a skew
partition described in sec.~\ref{sec:tmb} and sec.~\ref{sec:mod_rules}.

The idea is to exploit the fact that the characters of  indecomposable
modules $\Delta_L(\mu)$, given in \cite{Ram95}, do not depend
on the semisimplicity of $B_L(N)$.
This and some properties of generalized
Schur functions, which are summarized in
\cite{Ram98}, can be used to prove  that 
\begin{equation}
  \label{eq:bf}
  \str_{V^{\otimes L}} D^{\otimes L} d = \sum_{\mu\in X_L} sc_\mu(D) \chi'_\mu(d),
\end{equation}
is true even for all $L$.
Here $\chi'_\mu(d)$ is the character of $d\in B_L(N)$ in the
representation provided by $\Delta_L(\mu)$.
The functions $sc_\mu(D)$ are
polynomials in the eigenvalues of $D\in \OSp(R|2S)$, which where 
introduced for the first time by Bars in \cite{Bars81} in an early attempt to
describe the supercharacters of $\OSp(R|2S)$.
They can be defined recursively as
\begin{equation}\label{eq:elchar}
  sc_n(D)= \oint \frac{d\,z}{2\pi i}\left(
  \frac{1}{z^{n+1}}-\frac{1}{z^{n-1}} \right)
  \frac{1}{\sdet(1-zD )},
\end{equation}
and
\begin{equation}
  \label{eq:onchar} 
  sc_\mu(D)= \frac{1}{2}\det\left( sc_{\mu_j-i-j+2}(D)+sc_{\mu_j+i-j}(D)\right).
\end{equation}

For $L\leq N$ the Brauer algebra $B_L(N)$  is
semisimple and $J=0$. Consequently, $\Delta_L(\mu)$
are irreducible and $X_L = X_L(S)$.
Because of the commuting actions of $\OSp(R|2S)$ and
$B_L(N)$ one can naturally consider $V^{\otimes L}$ as a $\OSp(R|2S)$-$
B_L(N)$-bimodule, with $\OSp(R|2S)$ acting from the left and $B_L(N)$
from the right.
Then, eq.~\eqref{eq:bf} can be understood as a consequence of the decomposition 
\begin{equation}
\label{eq:decomp}
  V^{\otimes L} \simeq \bigoplus_{\mu \in X_L} G(\mu) \otimes
  \Delta_L(\mu), \quad L\leq N
\end{equation}
with $sc_\mu$ being actual characters of
tensor irreducible modules $G(\mu)$ as shown in \cite{Ram98}.

For $R,S$ such that $L>N$,
the polynomials $sc_\mu$ cannot generally be interpreted as the
character of some $\OSp(R|2S)$ representation.
As we have seen in sec.~\ref{sec:1749}, $\str_{V^{\otimes
L}}D^{\otimes L}d$ can be brought to the form $N^h \prod_m \str_V D^m$ and
eq.~\eqref{eq:bf} is not more then a simple equality between
two polynomials in eigenvalues of $D$.
% From eq.~(\ref{eq:isdecomp},\ref{eq:sidecomp}) the character of the
% representation $\OSp(R|2S)\times B_L(N)$ on
% $V^{\otimes L}$ is 
% \begin{equation}
%   \label{eq:ncf}
%   \str_{V^{\otimes L}}D^{\otimes L}d = \underset{\lambda\in
%     Y_L(S)}{\sum_{\mu\in X_L(S)}}g(\mu,\lambda)\sch_\lambda(D)
%   \chi_\mu(d),
% \end{equation}
% where  $g(\mu,\lambda)$ denotes the number of
%  irreducible components $G(\lambda)$ in $\mathcal{I}G(\mu)$ and
%  also the number of irreducible
%  components $B(\mu)$ in $\mathcal{I}B(\lambda)$.
Moreover, the two eqs.~(\ref{eq:cd},\ref{eq:bf}) are still compatible,
even if there are much more elements in $X_L$ then in $X_L(S)$. This
is possible because $sc_\mu$ are not functionally independent when $L>N$.
Then, for $\mu\notin Y_L(S)$ the polynomials
$sc_\mu$ can be written in terms of functionally independent $sc_\lambda$ with
$\lambda\in Y_L(S)$ by means of modification rules for characters
\begin{equation}
  \label{eq:mod_r}
  sc_\mu = \sum_{\lambda \in Y_L(S)}a(\mu,\lambda)sc_\lambda
\end{equation}
given in \cite{King87} and discussed in details in sec.~\ref{sec:mod_rules}.
% We show there that if $\lambda$ is a
% partition then $a(\mu,\lambda)\neq 0\Rightarrow \lambda\leq \mu$.
% It is convenient to write $\lambda^*\leq \mu$ if
% $a(\mu,\lambda^*)\neq0$.

The fundamental eq.~\eqref{eq:bf} is useful for small
widths $L$, when it is possible to compute the number $b'(\mu,\nu)$ of
irreducible components $B_L(\nu)$ in $\Delta_L(\mu)$ either by repeated
applications Frobenius reciprocity, as explained in \cite{Martin06}, or by
numerically
diagonalizing the transfer matrix of sec.~\eqref{sec:spin} in the
adjoint representation of $B_L(N)$ and detecting the ``accidental
degeneracies'' in its spectrum.
Indeed, from the explicit definition
\eqref{e:gen_irr} it is clear how to restrict the adjoint transfer matrix
to indecomposable modules $\Delta_L(\mu)$.
After we described in details 
the action of generators on the basis of $\Delta_L(\mu)$ in
sec.~\ref{sec:br_alg}, the algorithm of a numerical diagonalization is
straightforward.

The information about the structure of $\Delta_L(\mu)$ and the
modification rules in eq.~\eqref{eq:mod_r} can now be used to bring
eq.~\eqref{eq:bf} to the form
\begin{equation}
\label{eq:cf1}
  \str_{V^{\otimes L}}D^{\otimes L}d = \underset{\lambda\in
    Y_L(S)}{\sum_{\mu,\nu \in  X_L}}
  a(\nu,\lambda)b'(\nu,\mu)sc_\lambda(D)\chi_\mu(d).
\end{equation}
% where we have used $b(\nu,\mu)\neq 0 \Rightarrow \nu\leq
% \mu$.
We see that $\mu\in X_L(S)$ iff~\footnote{Although $sc_\lambda$, $\lambda\in
H_L(S)$
  are not irreducible $\osp(R|2S)$ characters, one can still 
use them as a basis for representing the character of any
representation.}
there is at least one $\lambda\in Y_L(S)$
such that $\sum_\nu
a(\nu,\lambda)b(\nu,\mu)\neq 0$.
To determine $g(\mu,\lambda)$ one has to decompose the factor of $\chi_\mu$
in eq.~\eqref{eq:cf1} as a sum of  $\OSp(R|2S)$ irreducible characters,
which are explicitly known, as far as we know, only for  $\OSp(3|2)$ and $\OSp(4|2)$.
Given the huge order of the set of weights $X_L$, it may seem that
calculations according to eq.~\eqref{eq:cf1}
are extremely cumbersome already for small $L$.
The simplifying point is that $a(\nu,\lambda)$ (or $ b'(\nu,\mu)$) is
non zero only if both weights are in the same
equivalence class of $Y_L(S)$ (or $X_L(S)$). 
The splitting of $Y_L(S)$ (or $X_L(S)$) into equivalence classes,
called \emph{blocks} and described in details 
in sec.~\ref{sec:tmb}, is with respect to an equivalence relation 
between irreducible components of indecomposable $\OSp(R|2S)$ (or
$B_L(N)$) modules.

% Virtually nothing
% is known about $\mathcal{S}B(\mu)$ and  $\mathcal{I}B(\mu)$.
% In a recent work \cite{Martin06} Martin \emph{et al} described the
% blocks of $B_L(N)$. 
% On the other side,  irreps and the blocks
% of $\osp(R|2S)$ where studied by Serganova in \cite{Serganova98}.
   
% We would like to describe in this section:
% first, the set of weights $X_L(S)$; second,
% a criterion of irreducibility for $\mathcal{I}G(\mu)$ and
% $\mathcal{I}B(\mu)$, and third, the composition factors
% of $\mathcal{I}G(\mu)$ and $\mathcal{I}B(\mu)$ when there are reducible.
% As our arguments are based only on the study of characters, we cannot
% say anything about the radical layer structure of indecomposables.

An important consequence of the fact that $\OSp(R|2S)$ supertrace
partition functions depends only on the $\OO(N)$ part of the spectrum
is the vanishing of the superdimension
$\sdim\mathcal{I}G(\mu) = 0$ for all indecomposable modules with $\mu
\notin X_L(0)$.
A more restrictive criterion for 
$\mathcal{I}G(\mu)$ supercharacters deriving from the full inclusion
sequence in eq.~\eqref{eq:filtr}
can be derived by taking a matrix $D$ with eigenvalues
$x_1=y_1$ and 
$x_i\neq y_j$ for $i=1\dots,r$, $j=1,\dots,S$.
Then, it can be seen from
eqs.~(\ref{eq:trp+},\ref{eq:loop_pf}) or
eqs.~(\ref{eq:elchar},\ref{eq:onchar},\ref{eq:bf}) that any $\OSp(R|2S)$
quasiperiodic
partition function
will also be an $\OSp(R-2|2S-2)$ quasiperiodic partition function.
As a consequence $ X_L(0)\subset X_L(1)\subset \dots \subset X_L(S)$
and the supercharacters
of $\mathcal{I}G(\mu)$ vanish when $\mu\in X_L(S)/X_L(S-1)$ and $D$
can be embedded in $\OSp(R-2|2S-2)$.

For $S=0$ the modules $V^{\otimes L}$ is semisimple. Therefore $\rad
B_L(L)\subset J(0)$. On the other hand, if $S$ is big enough $J(S)=0$.
It could be interesting to understand the relation between the sequence
$J(0)\supset J(1)\supset \dots\supset J(S)=0$ and the
cohomology of the radical $\rad B_L(N)\supset \rad^2 B_L(N)\supset \dots\supset 0$.

Observe  that a filtration similar to that of $X_L(S)$
is available on $Y_L(S)$ by the degree of
atypicality of  its elements.
In fact, we explain in sec.~\ref{sec:tmb} of the appendix how the weights in a block of
$X_L(S)$ or $Y_L(S)$  can be organized by the number of \emph{removable
balanced continuous border strips} in the corresponding Young
tableau.
This number can be interpreted as the degree of atypicality when the
corresponding partition represents an $\OSp(R|2S)$ weight.

\subsection{$\OO(2)$ spin model}

\begin{figure}
  \psfrag{0}{$\mu_0$}
  \psfrag{*}{$\mu_{0^*}$}
  \psfrag{1}{$\mu_1$}
  \psfrag{2}{$\mu_2$}
\psfrag{3}{$\mu_3$}
  \centering
  \includegraphics{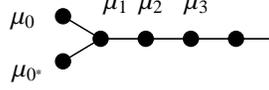}
  \caption{Edges in the graph represents nonzero elements of the
    fusion matrix $C$.}
\label{fig:fusion}
\end{figure}

Let $V$ be a two dimensional vector space endowed with an action of
$\OO(2,\mathbb{R})$. 
The action of $\OO(2)$ on the tensor space
$V^{\otimes L}$ is
\begin{equation*}
  \label{eq:o2act}
 D \cdot  V^{\otimes L} = \underbrace{D V\otimes \cdots \otimes
   DV}_L,\quad D\in \OO(2).
\end{equation*}
$B_L(2)$ acts according to the following definition of generators $E_i,P_i$ 
\begin{align*}
  %\label{eq:b2act}
  E_i &= \underbrace{1_2 \otimes \dots \otimes 1_2}_{i-1}
  \otimes \,E \otimes 1_2\otimes 
  \dots \otimes 1_2\\
  P_i &= \underbrace{1_2 \otimes \dots \otimes 1_2}_{i-1} \otimes \,P
  \otimes 1_2\otimes \dots \otimes 1_2,\\
\end{align*}
where $P,E$ have the following representation on $V^{\otimes 2}$
\begin{equation}
  \label{eq:pe2}
  P=\begin{pmatrix}
    1& 0& 0& 0\\
    0& 0& 1& 0\\
    0& 1& 0& 0\\
    0& 0& 0& 1
    \end{pmatrix},\quad
    E=\begin{pmatrix}
    0& 0& 0& 0\\
    0& 1& 1& 0\\
    0& 1& 1& 0\\
    0& 0& 0& 0
    \end{pmatrix}.
\end{equation}
(note that $E$ differs in some essential way from the projection operator onto
the singlet representation in the usual $SU(2)$ basis). The decomposition of
$V^{\otimes L}$ as a $\OO(2)$-$B_L(2)$-bimodule is simply
\begin{equation}
  \label{eq:o2decomp}
  V^{\otimes L}\simeq \bigoplus_{\mu \in X_L(0)}G(\mu)\otimes B_L(\mu).
\end{equation}
Here
\begin{equation*}
 X_L(0) \text{~is composed of partitions~} \mu_0=\emptyset, \mu_{0^*}=1^2, \text{~and~}
\mu_k=k,k \geq 1.
\end{equation*}
The tensor representations $G(\mu_k)$ are irreducible with dimensions 
$\dim G(\mu_{k})=1, k=0,0^*$ and $\dim G(\mu_k)=2, k\geq 2$.
The representation $G(\mu_0)$ is the trivial one and $G(\mu_{0^*})$ is the
associate one dimensional $\det D$ representation.
At the restriction to $\SO(2)\simeq \UU(1)$ the representations
$G(\mu_0)$ and $G(\mu_{0^*})$ become equivalent, while $G(\mu_k),k\geq
1$ splits into two nonequivalent one dimensional representations $
e^{\pm i k \phi }$ where $\phi$ is the $\UU(1)$ angle.
The $B_L(2)$ representations $B_L(\mu)$ are irreducible as well and
are constructed by acting with $B_L(2)$ on the tensor module in
eq.~\eqref{eq:gend}.
The dimensions of simple modules $B_L(\mu_k)$ are easily computed 
by looking at the first row of $C^L$, where $C$ is the fusion matrix
\begin{equation}
  \label{eq:adjmat}
  G(\mu_k)\otimes V \simeq \bigoplus C_{k,l}G(\mu_l)
\end{equation}
described by a $D_L$ type Dynkin diagram with labeling of the nodes
shown in fig.~\ref{fig:fusion}.
It is not hard to solve the recurrence relations satisfied by 
$d(L,k):=\dim B_L(\mu_k)$
\begin{align*}
  d(L+1,0)&=d(L+1,0^*)=d(L,1),\\
  d(L+1,1)&=2d(L,0)+d(L,2),\\
  d(L+1,k)&= d(L,k+1)+d(L,k-1), \quad k\geq 2
\end{align*}
and get that $d(L,k)=C_L^{[L/2]+k}$ except the case when $L$ is even
and $k=0$ for which $d(L,0)=d(L,0^*)=C_L^{L/2}/2$.
These are, as expected, the number of eigenvalues of the transfer
matrix for the 6 vertex model in the sector of spin $s_z=k/2$.

We are interested in giving an algebraic description of the
6 vertex model transfer matrix algebra.
In other words we want to identify the
Brauer algebra annihilator $J:=J(0)$ of  $V^{\otimes L}$
and  carry out the quotient $B_L(2)/J$.

All the $B_L(2)$ weights $\lambda\in X_L$ such that $\mu_k\subset \lambda$
satisfy either $\nu_0=1^3 \subseteq \lambda$ or $\nu_1=21\subseteq
\lambda$.
Thus, it is enough to consider $L=3$ and impose the vanishing of
the double sided ideal of the word $W_0,W_1\in B_L(2)$ projecting onto
$\Delta_3(\nu_i), i=0,1$.
As explained in the beginning of the previous section $W_i =\mathcal{T}_3
e_{\nu_i}$, where $\mathcal{T}_3$ extracts all the traces from $V^{\otimes 3}$
and $e_{\nu_i}$ are the Young symmetrizers corresponding to $\nu_i$.

The projector $\mathcal{T}_3$ can be found by looking at the form of
an arbitrary tensor $G_{ijk}$ after extracting all of its traces  
\begin{equation*}
G_{ijk}-\frac{\delta_{ij}}{4} \biggl( 3 G_{\cdot\cdot k} -
G_{k\cdot\cdot}-G_{\cdot k\cdot} \biggr) 
- \frac{\delta_{ik}}{4} \biggl( 3 G_{\cdot j\cdot} -
G_{j\cdot\cdot}-G_{\cdot\cdot j} \biggr)  -\frac{\delta_{jk}}{4}
\biggl( 3 G_{i\cdot \cdot} - G_{\cdot i\cdot}-G_{\cdot\cdot i}
\biggr),
\end{equation*}
which gives
\begin{align*}
\mathcal{T}_3 = & 1 -  \frac{1}{4} \biggl( 3 E_1 - E_2 E_1 - P_2 E_1
\biggr) -  \frac{1}{4} \biggl( 3 P_1 E_2 P_1 - E_1 P_2 -E_2 P_1
\biggr)  -  \frac{1}{4} \biggl( 3E_2 - E_1 E_2 - P_1 E_2 \biggr)
\end{align*}
and clearly $\mathcal{T}_3E_1 = \mathcal{T}_3 E_2 = \mathcal{T}_3 P_1
E_2 P_1 = 0$.

The Young symmetrizer $e_{\nu_0}$ is
\begin{equation*}
e_{\nu_0} = \frac{1}{6}\biggl( 1 + P_1 P_2 + P_2 P_1 - P_1 -P_2 -
P_1 P_2 P_1 \biggr)
\end{equation*}
and $e_{\nu_1}=e_{T_1}+e_{T_2}$, where
\begin{align*}
 e_{T_1} &= \frac{1}{3}\biggl( 1 - P_1P_2P_1\biggr)\biggl( 1+ P_1\biggr)\\
 e_{T_2} &= \frac{1}{3}\biggl( 1 - P_1 \biggr)\biggl( 1+ P_1P_2P_1\biggr)
\end{align*}
are the projectors onto the \emph{standard} Young tableau $T_1=[12,3]$
and $T_2=[13,2]$.
The two orthogonal projectors $e_{T_1}$ and $e_{T_2}$ are independent only
if we restrict to the right $B_L(2)$ action.
In fact, the left ideal of the word $W_1=0$
is the same as the double sided ideal of the word $\mathcal{T}_3 e_{T_1}=0$.

The condition $W_0 = 0$ gives the following restriction
\begin{equation}
\label{eq:st1}
1 + P_1 P_2 + P_2 P_1 = P_1 + P_2 + P_1 P_2 P_1
\end{equation}
on generators $P_1,P_2$.
Putting $P_i = 1-Q_i, i=1,2$ one can see that eq.~\eqref{eq:st1}
implies that $Q_i$ are
Temperley Lieb operators with $Q_i^2= 2 Q_i$.
There are no more restrictions that can be drawn from the conditions
$W_0=0$, because $W_0$ is a one dimensional projector.

Before exploring the next vanishing condition let us revise the
the defining relations of $B_L(2)$ given in eq.~\eqref{e:def_rel}  
\begin{align}\label{e:first_step0}
E_i P_i = P_i E_i = E_i \quad &\Rightarrow \quad Q_i E_i = E_i Q_i = 0 \\ \label{e:first_step}
P_i P_{i+1} E_i = E_{i+1}E_i \quad &\Rightarrow \quad Q_i Q_{i+1} E_i
= E_{i+1}E_i + Q_{i+1} E_i - E_i \\ \label{e:first_step2}
E_{i+1}P_i P_{i+1} = E_{i+1}E_i \quad &\Rightarrow
\quad E_{i+1}Q_i Q_{i+1} = E_{i+1}E_i +E_{i+1}Q_i - E_{i+1},
\end{align}
which imply 
\begin{align}\label{e:first_step20}
E_i Q_{i\pm 1}E_i &= E_i\\ \label{e:first_step1}
Q_i Q_{i+1} E_i &= Q_i E_{i+1} E_i
\end{align}
Observe that although the algebra has now two Temperley Lieb operators
their role is not symmetric yet at this stage.

Next, the condition $\mathcal{T}_3 e_{T_1}=0$ implies
\begin{equation*}
1+ P_1 - P_1 P_2 P_1 - P_1P_2 = 2E_1 +E_2   -E_1 E_2 - 2E_2 E_1 -
E_1 P_2 +E_2 P_1
\end{equation*}
%
% \begin{equation*}
% 1+P_1P_2P_1 -P_1 - P_2P_1 =  2P_1 E_2 P_1 + E_2  - 2E_2 P_1 - P_1 E_2
% + E_2E_1 - P_2 E_1  
% \end{equation*}
%
which after inserting $P_i=1 - Q_i$ with the help of
eqs.~(\ref{e:first_step},\ref{e:first_step1}) becomes
\begin{equation}
\label{eq:step4} 
Q_1 + 2Q_2 - Q_2 Q_1 - 2 Q_1 Q_2 = E_1 + 2 E_2 - 2 E_2 E_1 - E_1 E_2 +
E_1 Q_2 - E_2 Q_1.
\end{equation}
Multiplying eq.~\eqref{eq:step4} by $Q_2$ on the right we get:
\begin{equation}
\label{e:first_step10}
E_2 Q_1 Q_2 = E_2E_1 + E_2 Q_1 - E_2 = E_1 Q_2 + Q_1 Q_2 - Q_2 = E_2E_1Q_2.
\end{equation}
which can be used to rewrite eq.~\eqref{eq:step4} as
\begin{equation}
\label{eq:finalf}
Q_1 Q_2 + Q_2 Q_1 - Q_1 - Q_2 = E_1 E_2 + E_2 E_1 - E_2 - E_1.
\end{equation}
Multiplying by $E_i, Q_i$ on the left and on the right of
eq.~\eqref{eq:finalf} and using only the relations between $Q_i$, the
relations between $E_i$ and eq.~\eqref{e:first_step0} one can get all
the eqs.~(\ref{e:first_step}--\ref{e:first_step10})
and also
 \begin{align*}
 Q_1E_2E_1 = Q_1Q_2 + Q_1 E_2 - Q_1 &= E_2E_1+Q_2E_1 - E_1=Q_1Q_2E_1\\
 Q_1 E_2 Q_1 &= Q_1
 \end{align*}
which establish a complete symmetry  between $E_i$ and $Q_i$.

The double sided ideal of $\mathcal{T}_3 e_{T_1}=0$ is composed of four linearly
independent words --- two generated by the left action and other two
generated by the right action of $B_L(2)$. 
It is useful to note that after taking the quotient of $B_3(2)$ we
are left with 10 independent words instead of 15, which is exactly
what we need for the 6 vertex local transfer matrix.

We give the following  abstract definition to the 6 vertex model
transfer matrix algebra $\mathcal{V}_L:=\End_{\OO(2)} V^{\otimes L}$ in
term of generators $E_i,Q_i$
\begin{align}
E_i^2 = 2E_i, \quad E_iE_{i\pm1}E_i = E_i, \quad& Q_i^2 = 2Q_i, \quad
Q_i Q_{i\pm1}Q_i = Q_i\\
E_i Q_i &= Q_i E_i = 0\\ \label{e:magic}
Q_i Q_{i+1} + Q_{i+1}Q_i - Q_i - Q_{i+1} &= E_iE_{i+1} + E_{i+1}E_i -
E_i - E_{i+1}\\ \notag
E_i E_{j}=E_j E_i,\quad Q_i Q_j =& Q_j Q_i,\quad E_iQ_j = Q_j E_i\\ \notag
|i-j|>1,\quad& i,j=1,\dots,L-1
\end{align}

The defining relations are symmetric under the transposition $T$, which
changes the multiplication order, under the reflection
$R:E_i\rightarrow E_{L-i}$ and under the involution $E^*=Q$.
Thus, if $W=0$ then $W^T=0$, $W^R = 0$ and $W^*=0$ is also true
for any word $W\in \mathcal{V}_L$. 

% Here are the basic\footnote{These properties are useful when searching
%   for a solution of the Yang-Baxter equation.} properties of
% $\mathcal{V}_L$ resulting from eq.~\eqref{e:magic} of the definition: 
% \begin{align}\label{e:properties}
% Q_i Q_{i+1}E_i &= Q_{i+1}E_i + E_{i+1}E_i - E_i\\ \notag
% Q_i Q_{i+1}E_i &= Q_{i} E_{i+1}E_i\\ \notag
% Q_i E_{i+1}Q_i &= Q_i \\ \notag
% Q_{i+1}E_{i}+ E_{i+1}E_{i}- E_{i} &= Q_i E_{i+1}+ Q_i Q_{i+1}- Q_i.
% \end{align}

Introducing the operators $S_i = 1- E_i -Q_i$, with the property
$S_i^2=1$, one can rewrite 
eq.~\eqref{e:first_step10} as
\begin{equation*}
Q_{i+1} = S_i E_{i+1}S_i.
\end{equation*}
Thus, one can eliminate all of the generators $Q_i, i\geq 2$ and leave
only $Q_1$ subject to satisfy
\begin{align}\notag
  Q_1 E_1 = E_1 Q_1 =0, &\quad Q_1^2 = 2Q_1\\\label{eq:grint}
  Q_1 E_2 Q_1 = Q_1, &\quad E_2 Q_1 E_2 = E_2 \\\notag
  Q_1 E_j = E_j Q_1, &\quad j\geq 3.
\end{align}

\begin{figure}
  \psfrag{q}{$Q_1$}
  \psfrag{=}{$=$}
  \psfrag{0}{$0$}
  \centering
  \includegraphics{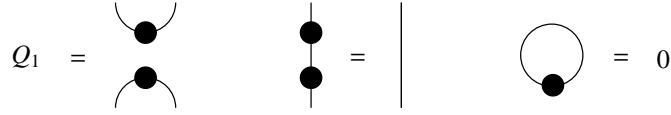}
  \caption{The generator $Q_1$ is represented as $E_1$
    with its horizontal edges marked by a blobbed.
    The conditions satisfied by the blob are represented on the right.} 
\label{fig:dn}
\end{figure}

Denote by $d_L$ the extension of the ordinary Temperley Lieb
algebra, generated by  $E_i$, with the additional generator $Q_1$
satisfying eqs.~\eqref{eq:grint}.
We see that $d_L$ and $\mathcal{V}_L$ are isomorphic algebras.
The graphical interpretation for the reduced words
(products of generators of minimum length) of $d_L$ and its relation
to the blob algebra and the Temperley Lieb algebra of type $D$ is 
discussed \cite{Green97}.
The generators $E_i$ are diagrammatically represented as usual, whereas
$Q_1$ is represented as $E_1$ with each of its horizontal edges marked
by an \emph{involutive} blob as shown in fig.~\ref{fig:dn}.
An unblobbed loop is identified with 2, while a blobbed loop with 0.
Thus, we see that $d_L$ is a subalgebra of the blob algebra composed
of all planar diagrams on 
$2L$ points with  an \emph{even} number of blobbed edges.
The dimension of $d_L$ is, as explained in
\cite{Green97}, half the dimension of the blob algebra, that
is $C^L_{2L}/2$.

There are several important consequences arising from the isomorphism
between $\mathcal{V}_L$ and $d_L$ from the point of view of
integrability.
First of all, we check that indeed the solution to 
the Yang-Baxter equation
\begin{equation}
\label{eq:yb}
R_1(u)R_2(u+v)R_1(v)=R_2(v)R_1(u+v)R_2(u)
\end{equation}
provided by the algebra $\mathcal{V}_3$ coincides with the well known
\emph{XXZ} spin chain $R$-matrix.

% \begin{equation}
%   \label{eq:xxzrmat}
%   R_{XXZ}(u) = \begin{pmatrix}
%     \sin(\lambda-u) & 0 & 0 & 0 \\
%     0 & \sin \lambda & \sin u & 0 \\
%     0& \sin u & \sin \lambda & 0\\
%     0& 0& 0& \sin(\lambda-u)
%     \end{pmatrix},\quad \Delta = - \cos \lambda
% \end{equation}
% if $Q=1-P$ and $E,P$ is of the form in eq.~\eqref{eq:pe2}.

For that, consider the ansatz $R(u) = I + f(u)Q + g(u) E$ and plug it
in eq.~\eqref{eq:yb}. 
Choosing as basis set in $\mathcal{V}_3$ the 10 words $1, E_1, E_2,
E_1E_2, E_2E_1, Q_1E_2, E_2Q_1, Q_1E_2E_1, E_1E_2Q_1$ we get two
independent functional equations
\begin{align}\label{eq:1yb}
E_1:\;F(f,g)-F(g,f)&=g'f-gf'+fg''-f''g\\ \label{eq:2yb}
E_2:\;F(f,g)+F(g,f)&=(f''g'+f'g'')(f+g),
\end{align}
where $F(f,g)=f''+f'-f+f'f''(2+f+g)$.
The primed functions are evaluated in $u$, the unprimed in
$u+v$ and the double primed in $v$.
All other words provide the same third equation, which is a
consequence of eqs.~(\ref{eq:1yb},\ref{eq:2yb}).
The solution to the system of eqs.~(\ref{eq:1yb},\ref{eq:2yb}) is
\begin{align}
  f(u)&=\frac{\sin \lambda - \sin u}{2 \sin (\lambda-u)}-
  \frac{1}{2}\\
  g(u)&=\frac{\sin\lambda+\sin u}{2 \sin(\lambda-u)}-\frac{1}{2},
\end{align}
with an arbitrary constant $\lambda$.
Taking $Q$ and $E$ in the representation provided by the
eq.~\eqref{eq:pe2} we find the famous \emph{XXZ} spin chain
$R$-matrix
 \begin{equation}
   \label{eq:xxzrmat}
   R_{XXZ}(u) = \begin{pmatrix}
     \sin(\lambda-u) & 0 & 0 & 0 \\
     0 & \sin \lambda & \sin u & 0 \\
     0& \sin u & \sin \lambda & 0\\
     0& 0& 0& \sin(\lambda-u)
     \end{pmatrix},\quad \Delta = - \cos \lambda
 \end{equation}
as expected.

Clearly, an integrable system in $\mathcal{V}_L$ has to be related to
an integrable system in $d_L$ because of the isomorphism of these two
algebras.
However, the ansatz $R(u)=1+g(u)E$ plugged into the eq.~\eqref{eq:yb}
gives only the isotropic point ($\Delta=\pm1$) solution
$g(u)=u/(1-u)$.
The only possibility to give a richer content to the integrability in
$d_L$ is by introducing nontrivial boundary conditions.
This means that the anisotropy of the \emph{XXZ} spin chain can be
generated by introducing nontrivial boundary conditions at the
isotropic points, an observation made earlier from a slightly different perspective in \cite{Levy}.

\subsection{$\OSp(4|2)$ spin model}

\label{heart}

The representation theory of the superalgebra $\osp(4|2)$ is
summarized in \cite{Germoni04}.
As we have already mentioned, all of $\osp(4|2)$ irreducible
characters have been
computed and indecomposable representations classified.
We give a brief reminder of these results in sec.\ref{sec:osp42}
and make some remarks, based on the general discussion in
sec.~\ref{sec:supergr}, on the difference between the
representation theory of the supergroup $\OSp(4|2)$ and its Lie
superalgebra.

The tensor space $V^{\otimes L}$, seen as a $\OSp(4|2)$ module, can be represented 
as a direct sum $V^{\otimes L}= V^{(0)}\oplus V^{(1)}$ of a part ``lifted'' from  $\OO(2)$
\begin{equation}
  \label{eq:v0}
  V^{(0)} = \bigoplus_{\lambda \in Y_L(0)} n_L^\lambda G(\lambda)
\end{equation}
and a projective part
\begin{equation}
  \label{eq:v1}
  V^{(1)} = \bigoplus_{\lambda\in Y_L(1)/Y_L(0)} n_L^{\lambda} \mathcal{P}G(\lambda),
\end{equation}
where $\mathcal{P}G(\lambda)$ is the projective cover of
$G(\lambda)$.
This decomposition can be proved by induction on $L$ using two facts:
\begin{itemize}
\item The tensor product between atypical irreducible
  representations with highest weights labeled by one row partitions (see bellow) and $V$
  decomposes to~\footnote{Modulo irreps labeled by  $\lambda\notin Y_L(0)$, this
decomposition provides the
    same fusion matrix as eq.~\ref{eq:adjmat}. Thus,  $Y_L(0)$
multiplicities in $V_{4|2}^{\otimes L}$ are the same as those in $V_{2|0}^{\otimes L}$.
Rather then a   coincidence, this is a direct manifestation of the
algebra inclusion $\End_{\OSp(4|2)}V_{4|2}^{\otimes L}\supset
    \End_{\OO(2)}V_{2|0}^{\otimes L}$ at the level of dimensions of
    irreps.}
  \begin{equation*}
    G(k) \otimes V \simeq G(k+1) \oplus G(k1) \oplus G(k-1).
  \end{equation*}
  This is proved by counting the dimensions on the right/left hand
  sides and, then, observing that $G(k1)$ is typical and $G(k\pm1)$,
  being in different blocks, cannot give rise to
  indecomposables.
\item The tensor product of a projective module with any other module
  is projective, thus, decomposing to a direct sum of projectives.

\end{itemize}

In the following we use the fundamental eqs.~(\ref{eq:bf},\ref{eq:cf1}) to
decompose $V^{\otimes L}$ as a $\mathsf{Z}=\End_{B_L(2)}V^{\otimes L}$ module
and verify the assumption that $\mathsf{Z}= \mathbb{Z}_2\times \osp(4|2)$ by
comparing the result to eqs.~(\ref{eq:v0},\ref{eq:v1}).

%The matrix $a(\mu,\lambda)$ is given in sec.~\ref{sec:tmb}.

The conditions of atypicality for a $\osp(4|2)$ weight $\lambda$ 
are given in
sec.~\ref{sec:osp42}.
In the partition notation we adopt, these are equivalent to
\begin{equation*}\label{at_cond_part}
 \lambda_1'=1 \; \text{or} \; \lambda_1+1 = \lambda_1' \; \text{or} \; \lambda_2 = \lambda_1'.
\end{equation*}
Typical weights satisfy none of atypicality conditions listed above.
Note that typical representations are irreducible, have vanishing
superdimension, and are simultaneously projective and injective.
This means they cannot be a constituents of any other 
$\osp(4|2)$ representations without being a direct summand.
One can say they are ``their own blocks''.

The supercharacters of associate $\OSp(4|2)$ irreps
$\lambda,\lambda^*$ satisfy $\sch_{\lambda^*}(D) = \sdet D \sch_\lambda(D)$.
For typical weights, the polynomials $sc_\lambda$    
give the right $\OSp(4|2)$ irreducible character.
Because of the modification rules, see sec.~\ref{sec:mod_rules}, 
it is possible to define a partition  $\lambda_\text{mod}$ such
that $sc_{\lambda_\text{mod}}(D)=\sdet(D) sc_\lambda(D)$.
Therefore, it is convenient to identify the associate weight
$\lambda^* = \varepsilon\times \lambda$ with the partition
$\lambda_\text{mod}$.
The Young tableau of $\lambda^*$ can be constructed by
replacing the orthogonal part of the Young tableau of
$\lambda$ by its associate, that is by putting $\big(\lambda^*\big)'_2=4-\lambda'_2$ and leaving all other
columns unchanged. For instance $(1^4)^*= 2^4$.
Exceptions are the typical weights $\lambda$ such that $\lambda'_1<4-\lambda'_2$.
The only such weights are $\lambda=1^3,21$ or $l1, l\geq 3$ and we put
$(1^3)^*=32^3,
(21)^*=3^221$ and $(l1)^*= l32, l\geq 3$.

The atypical $\OSp(4|2)$ weights can be labeled by two
integers $k$ and $l$, where $k$ denotes the isomorphism
class, also called  block.

In the partition notation, the block $k=0$ is composed of weights
$\lambda_{0,0}=\emptyset, \lambda_{0,1}=(1^2)^* := 3^22^2$
and $\lambda_{0,l}=(l1^l)^* := l2^21^{l-1}, l\geq 2$.
The associate block $k=0^*$ is composed of weights
$\lambda_{0^*,0}=1^2$ and $\lambda_{0^*,1} =(\emptyset)^*:= 3^4,
\lambda_{0^*,l}=\lambda_{0,l}^*=l1^l, l\geq 2$.

The self associate blocks $k\geq 1$ are composed of weights
$\lambda_{k,0}=k, \lambda_{k,1}=k^*:= k3^2$ ($3^32$ for $k=1$ and
$3^31$ for $k=2$),
$\lambda_{k,l}=\lambda_{k,l}^*=kl1^{l-2}$ for $2\leq l \leq k$ and
$\lambda_{k,l}=\lambda_{k,l}^*=l(k+1)1^{l-1}$ for $l\geq k+1$.

With the given notation for associate weights one can check with the
help of \cite{Germoni04} the following
decomposition of polynomials $sc_{\lambda_{k,l}}$ as a sum of $\OSp(4|2)$
supercharacters $\sch$
\begin{align}
  \label{eq:bk0}
  sc_{\lambda_{k,0}} &= \sch_{\lambda_{k,0}}, \;
  sc_{\lambda_{k,1}}=-\sch_{\lambda_{k,3}}+ \sch_{\lambda_{k,1}},\;
sc_{\lambda_{k,2}} = \sch_{\lambda_{k,2}}+ \sch_{\lambda_{k,0}},\;\\ \notag
sc_{\lambda_{k,l}} &= \sch_{\lambda_{k,l}}+ \sch_{\lambda_{k,l-1}}+
(-1)^{l-1}\sch_{\lambda_{k,1}}, \;  l \geq 2.
\end{align}
This is done in two steps. First one show that eqs.~\eqref{eq:bk0} hold for
a supermatrix $D$ with $\sdet D=1$.\footnote{To compare with \cite{Germoni04}
one has to take the
  eigenvalues of $D$ of the form $e^{\pm
  \epsilon_1},e^{\pm \epsilon_2\pm \epsilon_3}$} At this step is
yet impossible to distinguish between  associate representations.
In order to do so, one has to explicitly construct the elements of
the enveloping Lie superalgebra connecting the maximal vectors of irreducible components
of indecomposable highest weight modules and, then, look at their symmetry
under the outer automorphism $\tau$. See sec.~\ref{sec:osp42} for details.

We have just listed all the elements of $Y_L(1)$.
Eq.~\eqref{eq:bk0} is a bijection between
$\sch_\lambda$ and $sc_\lambda$.
As a consequence, $\OSp(4|2)$ and $B_L(2)$ weights can be labeled by
the same set $Y_L(1)=
X_L(1)$ in the partition notation we have adopted.
This is supporting the assumption that there is some sort of exact
equivalence between the category of $\OSp(4|2)$ and $B_L(2)$ modules
on $V^{\otimes L}$.
Bellow all the weights are partitions and, to avoid confusion,
we write $\lambda\in Y_L(1)$ if $\lambda$ is considered as a
$\OSp(4|2)$ weight and $\lambda\in X_L(1)$ if it is considered as a
$B_L(2)$ weight.

Let us show that the terms in eq.~\eqref{eq:bf} with
$\lambda\notin Y_L(1)$ do not actually contribute to
$\str_{V^{\otimes L}} D^{\otimes L}d$.
First note that if $\chi_\lambda$ cancels out from 
eq.~\eqref{eq:bf} then certainly $\delta_L(\lambda)$ in
eq.~\eqref{eq:gend} is a trivial module.
Therefore any module $\delta_L(\nu)$ will also be trivial if
$\lambda \subset \nu$.
Second, if $\chi_\lambda$ does not contribute to
eq.~\eqref{eq:bf} when $\lambda\vdash L$ then it does not contribute
to it for any $L$.
Thus, it is enough to prove for every $k$ that the
weights just greater (by inclusion) then $\lambda_{k,l}$
do not contribute to eq.~\eqref{eq:bf} when the are allowed for the
first time to appear.

Let $\lambda\in Y_L(1)$ be a typical (associate) weight.
Then, as we show in sec.~\ref{sec:tmb}, $\lambda\in X_L(1)$ is
a minimal partition (with respect to the inclusion in its block).
There will be a unique weight $\nu\notin Y_L(1)$ just greater then $\lambda$ and,
\emph{a priori}, $sc_\nu$ can modify to $\pm sc_\lambda$.
It is proved by induction in sec.~\ref{sec:mod_rules} 
that a positive sign would imply 
atypicality conditions on $\lambda$ and, thus, $sc_\nu=-sc_\lambda$.
Moreover, from \cite{Martin06} we know that $\Delta_L(\lambda)$ has
one composition factor $B_L(\nu)$.
Taking $L=|\nu|$, we see that the contribution to  eq.~\eqref{eq:bf}
of $\chi_\nu$ from $\Delta_L(\lambda)$ cancels out with the one from
$\Delta_L(\nu)$.

\begin{figure}
  \psfrag{a}{$\lambda_{k,0}$}
  \psfrag{b}{$\lambda_{k,2}$}
  \psfrag{c}{$\lambda_{k,3}$}
  \psfrag{d}{$\lambda_{k,4}$}
  \psfrag{e}{$\lambda_{k,5}$}
  \psfrag{f}{$\lambda_{k,1}$}
  \psfrag{g}{$\nu_k$}
  \centering
  \includegraphics{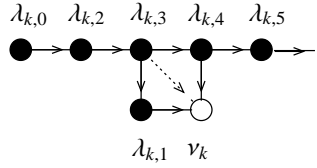}
  \caption{Two weights $\lambda_{k,l}$ and $\lambda_{k,l'\neq l}$ are
    connected by a continuous line iff $\lambda_{k,l}\subset
    \lambda_{k,l'}$ and there is no other weight between them. 
    The weight $\lambda_{k,l}$ is connected to $\lambda_{k,l'}$ by a
    dotted arrow iff  $\Delta_L(\lambda_{k,l})$ has 
    an irreducible component $B(\lambda_{k,l'})$. Its multiplicity is
    always  one.
  } 
\label{fig:einfty}
\end{figure}

Before proceeding to nontrivial blocks we need
to know the number of irreducible components $B_L(\lambda_{k,l'})$ in
$\Delta_L(\lambda_{k,l})$.
According to \cite{Martin06}, the graph representing the
partial ordering (by inclusion) of weights in a block $k$ determines
the \emph{required information} about the content of modules $\Delta_L(\lambda_{k,l})$.
The ordering graph is represented in fig.~\ref{fig:einfty}.

Now, let $\lambda_{k,l}\in Y_L(1)$ be an atypical (associate) weight.
Then, any weight $\nu\notin X_L(1)$ such that $\lambda_{k,l}\subset
\nu$ satisfies $\nu_k\subseteq \nu$, with $\nu_k$ represented by a
white dot in fig.~\ref{fig:einfty}. 
The explicit form of $\nu_k$ is $\nu_0=432^21,
\nu_{0^*}=43^31, \nu_1=43^221, \nu_2=43^21^2,\nu_3=4^231^2$ and
$\nu_k=k431, k\geq 4$.
Next, one can check with the help of modification rules that
\begin{equation}
  \label{eq:tempref}
sc_{\nu_k}+sc_{\lambda_{k,1}}+sc_{\lambda_{k,3}}+sc_{\lambda_{k,4}}=
0
\end{equation}
vanishes identically.
Further, from fig.~\ref{fig:einfty} each of the modules
$\Delta_L(\lambda_{k,l}), l=1,3,4$ has a single irreducible component
$B_L(\nu_k)$.
Finally, taking $L=|\nu_k|$ one can see from eq.~\eqref{eq:tempref}
that the contribution of $\chi_{\nu_k}$ to eq.~\eqref{eq:bf} cancels
out.

Let us introduce the compact notations $B_{k,l}:=
B_L(\lambda_{k,l})$ and $G_{k,l}:= G(\lambda_{k,l})$.
Then, putting together eq.~\eqref{eq:bk0} and fig.~\ref{fig:einfty} we get
from eq.~\eqref{eq:bf} the following content of indecomposable
modules $\mathcal{I}G_{k,l}$ appearing
in eq.~\eqref{eq:cd}
\begin{align}\notag
  B_{k,0}& \quad& B_{k,2}& \quad&
  B_{k,1}& \quad& B_{k,3}& \quad& B_{k,l+1}&
  \\[7.5pt] \notag
  & \quad& G_{k,0}& \quad& G_{k,1}&
  \quad& G_{k,2}& \quad& G_{k,l}\;& \\
  \label{eq:rez1}
  G_{k,0}& \quad& G_{k,2}& \quad& G_{k,2}&
  \quad& G_{k,0}\,G_{k,1}&\,G_{k,3} \quad&
  G_{k,l-1}\;G&_{k,l+1},  \\ \notag
  & \quad& G_{k,0}& \quad&  G_{k,1}& \quad&
  G_{k,2}& \quad& G_{k,l}\;&
\end{align}
where $l=3,\dots,m$ and $\lambda_{k,m}\vdash L$.
The indecomposable modules $\mathcal{I}G_{k,l}$ are
represented below $B_{k,l}$ and it should be understood that
they get ``paired up'' in the decomposition of $V^{\otimes L}$ as a
$\OSp(4|2)\times B_L(2)$ bimodule.\footnote{In pedantic terms, the
pairing $B_{k,l},\mathcal{I}G_{k,l}$ can be
represented by the functor $B_{k,l}\rightarrow V^{\otimes
  L}\otimes_{B_L(2)}B_{k,l}$ sending $B_L(2)$ left modules to
$\mathbb{Z}_2\times \osp(4|2)$ left modules.}
Alternative, maybe more intuitive physically, representations of the blocks will be given in the next paper.

The structure of modules $\mathcal{I}G_{k,l}$ is in perfect
agreement with eq.~(\ref{eq:v0},\ref{eq:v1}).
We recognize in the first term $\mathcal{I}G_{k,0}=G_{k,0}$ of
eq.~\eqref{eq:rez1} the contribution to $V^{(0)}$,
while the rest of the terms are exactly the projective modules
appearing in $V^{(1)}$, that is 
$\mathcal{I}G_{k,2}=\mathcal{P}G_{k,0},
\mathcal{I}G_{k,1}=\mathcal{P}G_{k,1}$ and
$\mathcal{I}G_{k,l}=\mathcal{P}G_{k,l-1}, l\geq 2$.

For typical $\lambda\in Y_L(1)$, the modules $\mathcal{I}G(\lambda)=G(\lambda)$
are irreducible and get paired up with $B_L(\lambda)$ in the
decomposition of $V^{\otimes L}$ as a $\OSp(4|2)$-$B_L(2)$-bimodule.

Observe that, as expected, only the modules $B_{k,0}$ (which
coincide with $B_L(\lambda_k)$ in eq.~\eqref{eq:o2decomp}) contribute to the
supertrace $\str_{V^{\otimes L}} d$.
Indeed, typical modules modules $G(\lambda)$ have superdimension 0.
The same is true for projective modules.
One can explicitly check from eq.~\eqref{eq:rez1} that $\sdim
\mathcal{P}G_{k,l}=0$ if we take into account that only
 $\osp(4|2)$ fermionic generators connect
irreducible components of indecomposable modules.
For instance, $\sdim \mathcal{P}G_{0,0} = \sdim
G_{0,0} - \sdim G_{2,0} + \sdim G_{0,0} =
1-2+1 = 0$.

As we have explained at the beginning of sec.~\ref{sec:gf}, the
degeneracies of the eigenvalues of the $\OSp(4|2)$ spin transfer
matrix are given by $\dim\mathcal{I}G(\lambda)$.
We compute them in app.~\ref{sec:osp42}.

Thus, in conclusion we see that 
$B_L(2)/J(1) = \End_{\mathbb{Z}_2 \times \osp(4|2)}V^{\otimes L}$ and, because $V^{\otimes
  L}$ is by definition a faithful $B_L(2)/J(1)$ module we also have 
$\mathbb{Z}_2\times \osp(4|2)=\End_{B_L(2)/J(1)}V^{\otimes L}$.
In other words, the two algebras $B_L(2)/J(1)$ and
$\mathbb{Z}_2\times \osp(4|2)$ are the full centralizers of each other on
$V^{\otimes L}$.

This results allows us to relate the decomposition of $V^{\otimes
  L}$ as a $\OSp(4|2)$
left module to the
decomposition of  $V^{\otimes
  L}$ as a $B_L(2)$ right module.

Collecting in a single indecomposable module $\mathcal{I}B_{k,l}$
all factors $B_{k,l'}$ in eq.~\eqref{eq:rez1} which correspond to (happen
to be above) an irreducible component $G_{k,l}$
we get~\footnote{Again this ``collecting'' can be represented by the
  functor $G_{k,l}\rightarrow \Hom_{\mathbb{Z}_2\times
    \osp(4|2)}(V^{\otimes L}, G_{k,l})$ sending
  $\mathbb{Z}_2\times\osp(4|2)$ left modules to $B_L(2)$ left modules.}
\begin{align}\notag
  G_{k,0}& \quad& G_{k,1}& \quad&
  G_{k,2}& \quad& G_{k,l-1}& \quad& G_{k,m-1}& \quad& G_{k,m}& 
  \\[7.5pt] \notag
  B_{k,2}& \quad& B_{k,1}& \quad& B_{k,3}&
  \quad& B_{k,l}\;& \quad& B_{k,m}\;& \\
  \label{eq:rez2}
  B_{k,0}\;B&_{k,3} \quad& B_{k,3}& \quad& B_{k,2}\,B_{k,1}&\,B_{k,4} \quad&
  B_{k,l-1}\,B&_{k,l+1} \quad& B_{k,m-1}& \quad& B_{k,m},& \\ \notag
  B_{k,2}& \quad& B_{k,1}& \quad&  B_{k,3}& \quad&
  B_{k,l}\;& \quad& B_{k,m}\;& \quad& &
\end{align}
where $l=4,\dots,m-1$ and the content of $\mathcal{I}B_{k,l}$ is
represented below $G_{k,l}$.

Apart the last irreducible module $\mathcal{I}B_{k,m}=B_{k,m}$, we
recognize in the terms of eq.~\eqref{eq:rez2} the projective
representations of the quiver $E_\infty$ in fig.~\ref{fig:einfty},
which describes the homomorphisms between the $B_L(2)$ tensor modules
$\delta_L(\lambda)$ realized on $V^{\otimes L}$.

\section{The hamiltonian limit}

It will turn out in our forthcoming analysis of conformal properties  to be easier to study numerically 
the hamiltonian
\begin{equation*}\label{e:gen_ham}
H_\Delta= -\frac{1+\Delta}{2}\sum^{L-1}_{i=1}\left( I + P_i\right)
-\frac{1-\Delta}{2}\sum_{i=1}^{L-1} E_i.
\end{equation*}
The expectation --- which we will confirm in great details
--- is that this hamiltonian will be in the same
universality class as the spin model we had started with.

The hamiltonian $H_\Delta$ is obviously local and has only nearest
neighbour interactions if the $E$'s and $P$'s are taken in the
spin representation provided by eq.~\eqref{e:osp_rep}.
However, this is no longer true if we think of $H_\Delta$ as an
element of the adjoint representation of $B_L(2)$.

The lowest eigenvalue of $H_\Delta$ belongs to the $B_L(2)$ irreducible
representation labeled by $\mu = L \mod 2$.

For generic $\Delta$ it is nondegenerate if $L$ is even and has
degeneracy $\dim V = 4S+2$ if $L$ is odd.
On the other hand, the highest eigenvalue belongs to the completely
antisymmetric representation labeled by $\mu = 1^L$.
In this representation the $P$'s act as -1  and the $E$'s as 0.

The hamiltonian $H_\Delta$ is determined up to an arbitrary additive
constant and multiplicative factor.
For numerical diagonalization it is convenient to fix the additive
constant such that the
maximal eigenvalue of $H_\Delta$ be zero.
The multiplicative factor is fixed by requiring
\begin{equation*}
 H_\Delta \Big\vert_{S=0} = H_{XXZ}+ \const %\frac{2+\Delta}{2}\bigl( L-1 \bigr),
\end{equation*}
with $I,E,P$ as in eq.~\eqref{e:osp_rep}, $J$ as in app.~\ref{sec:superalg} and
$H_{XXZ}$ being the \emph{XXZ} spin chain hamiltonian
in its usual form
\begin{equation*}\label{e:xxz_ham}
H_{XXZ} = -\frac{1}{2}\sum_{i=1}^{L-1} \left( \sigma^x_i \otimes 
\sigma^x_{i+1}+ \sigma^y_i \otimes \sigma^y_{i+1}+\Delta\,\sigma^z_i 
\otimes \sigma^z_{i+1} \right).
\end{equation*}

The fact that the eigenvalues of the 6 vertex model appear as a 
subset 
of the eigenvalues of the transfer matrix for the $\OSp(2S+2|2S)$ 
model and  $S\geq 1$ carries over to a similar result for the hamiltonians. 
The velocity of sound for the massless excitations 
can thus be derived from its value for the \emph{XXZ} subset, which is well 
known from \cite{ham86} to be 
\begin{equation}
  \label{eq:fvel}
    v_{s}={\pi\sin\lambda\over\lambda},\quad \Delta=-\cos\lambda.
\end{equation}

The hamiltonian $H_{\Delta}$ is diagonalized numerically in the 
adjoint representation of the Brauer algebra  by studying its action on the diagrams just 
like for the transfer matrices.
Next, once the structure of indecomposable modules $\Delta_L(\mu)$ is
known,  eq.~\eqref{eq:bf} can be used as explained in
sec.~\ref{sec:gf} to
select the part of the spectrum which does indeed appear for a fixed
$S$ spin model.

However, in the two special cases  $\Delta=\pm 1$ the hamiltonian
$H_\Delta$ greatly simplifies.
In the following two subsections we discuss the behaviour of the
spectrum of $H_\Delta$ in the two limits $\Delta \to \pm 1^{\mp}$.

\subsection{The limit $\Delta = 1$}

When $\Delta = 1$ the Temperley Lieb
operators $E_i$ do not contribute to $H_\Delta$ and, thus, the
hamiltonian is no longer a generic element of the Brauer algebra
$B_L(2)$, but belongs instead to the subalgebra
$\mathbb{C}\Sym (L)\subset B_L(2)$.
This will translate to additional degeneracies in the spectrum of
$H_\Delta$ at the point $\Delta=1$ compared to other points in the
range $-1 \leq\Delta <1$.

Hamiltonians of type $-\sum P_i$, with $P$'s in the representation
provided by eq.~\eqref{e:osp_rep}, are integrable and have been
studied in \cite{suth75} and \cite{hubert99}.
Although the continuum limit of such spin chains is a gapless field
theory, it fails to be
conformal, because excitations have a  $L^{-2}$ scaling law in the thermodynamic limit.
This can readily be seen from the vanishing of the sound velocity in eq.~\eqref{eq:fvel}.
We will not enter into the details here, but just mention that the
different systems of Bethe ansatz equations are indexed by
$(2S+2,2S)$-hook shape partitions $\lambda\vdash L$.
This is exactly the label of irreducible representations of
the group algebra $\mathbb{C}\Sym(L)$ realizing in the centralizer of the spin
chain $V^{\otimes L}$, with $V$ being the fundamental representation of $\SU(2S+2|2S)$.
We see that the symmetry of our spin model $\OSp(2S+2|2S)$
jumps to $\SU(2S+2|2S)$ at the point $\Delta=1$.

The additional degeneracies in the spectrum of the $\OSp(2S+2|2S)$ spin model at the
point $\Delta=1$ can be understood by looking at the decomposition 
of
$B_L(2)$ modules $\Delta_L(\mu)$, into a direct sum of
$\mathbb{C}\Sym(L)$ irreducible modules $S(\lambda)$.
Let $\mu\vdash L-2k$ and  $\lambda\vdash L$, then  it was shown in
\cite{hw90} that the multiplicity of $S(\lambda)$ in the
decomposition of $\Delta_L(\mu)$ is
\begin{equation}
  \label{eq:m}
  m(\mu,\lambda)= \underset{\eta \text{ even}}{\sum_{\eta\vdash 2k}} c^\lambda_{\mu\eta},
\end{equation}
where $c^\lambda_{\mu\eta}$ are Littlewood-Richardson coefficients.
Alternatively, $m(\mu,\lambda)$ is the number of
tensors of rank $L-2k$, with index symmetry of some fixed standard
Young supertableau of shape $\mu$, that can be obtained from a
tensor of rank $L$, with index symmetry of some standard Young
supertableau of shape $\lambda$, by contracting $2k$ indices
in all the possible ways.  

One can apply eq.~\eqref{eq:m} to understand the
degeneracy of the lowest level of $H_\Delta$ at $\Delta=1$.
First, observe that  $-\sum P_i$ is minimized in
the sector $\lambda = L$  (where $P$'s acts as 1).
The \emph{only} $\mu$ such that $m(\mu,\lambda)\neq 0$ are one row partitions.
Thus, the lowest eigenvalues of $H_\Delta$ restricted to
$\Delta_L(L-2k)$ for $k=0,\dots,[L/2]$  become all degenerate at $\Delta=1$.

Arguments of this kind can be used to derive information about
the critical exponents of the spin model in the limit $\Delta \to
1^-.$

\subsection{The limit $\Delta= -1$}

The same reasoning can be applied to the point $\Delta = -1$.
At this point, the hamiltonian $H_\Delta$ belongs to the Temperley Lieb
subalgebra  $T_L(1) \subset B_L(2)$
and the model can be considered as a spin chain $(V\otimes
\bar{V})^{\otimes \frac{L}{2}}$
where $V,\bar{V}$ are the fundamental representation of $\SU(2S+2|2S)$ and its conjugate. 
Additional degeneracies can be understood by looking at the
decomposition of $B_L(2)$ modules $\Delta_L(\mu)$ as a direct sum of
standard irreducible $T_L(1)$ modules $D_L(j)$.

Let us compute the multiplicity $n_L(\mu, j)$ of irreducible modules
$D_L(j)$ in the decomposition of $\Delta_L(\mu)$ with $\mu \vdash L-2k$.

As explained in sec.~\ref{sec:br_alg}, $\Delta_L(\mu)$ has a natural
basis composed of all possible pairings $p \otimes v_i$ of partial diagrams
$p$ with $m=L-2k$ free points and basis vectors $v_1,\dots, v_{f_\mu}$ of $S(\mu)$. 
We say that a horizontal line of a partial diagram $p$ is \emph{intersected} either if it
intersects another horizontal line or if there is a free point in $p$ between
the two ends of the horizontal line.
Let us associate to each partial diagram $p$ the
number of intersected horizontal lines $l$ in $p$.
It is not hard to see that the span on the basis vectors $p\otimes v_i$,
with $p$'s having at most $l$ horizontal intersected lines, is a $T_L(1)$ submodule
in $\Delta_L(\mu)$.
If we denote this submodule by $\Delta^l_L(\mu)$ there is an obvious
filtration  
$\Delta_L(\mu) = \Delta^k_L(\mu) \supset \dots \supset
\Delta^0_L(\mu)\supset \Delta^{-1}_L(\mu) = 0$ of $\Delta_L(\mu)$.

Consider the natural action of $T_L(1)$ on the quotient modules
$Q^l_L(\mu) = \Delta^l_L(\mu)/\Delta^{l-1}_L(\mu)$.
Observe that the action of $T_L(1)$ changes the labeling $\pi\in \Sym(m)$ of
free points in a labeled graph $p\otimes \pi$ if and only if it also reduces the number of horizontal
intersected lines.
Therefore, $Q^l_L(\mu)$ is isomorphic to a direct sum of $f_\mu$
modules $ Q^l_L(m)$.
Obviously $Q^0_L(m)\simeq D_L(m)$ and, therefore, we get 
$n_L(\mu,j)=0$ for $j< m$, $n_L(\mu,j) =
f_\mu n_L(m,j)$ for $ m \leq  j\leq  L$ and finally $n_L(m,m)=1$.

Thus, our problem effectively reduces to understanding the
action of $T_L(1)$ on the module $\Delta_L(m)$, which is composed of partial diagrams $p$ on
$L$ points with $m$ unlabeled free points.

At a closer look, one can see that the action of $T_L(1)$ on
partial diagrams keeps the reciprocal configuration of intersected lines and free
points intact.
In other words, if $\psi$ is a map that
eliminates all the nonintersected horizontal lines from a partial
diagram and acts
as identity otherwise,  then $\psi$ defines an invariant of $T_L(1)$,
that is 
\begin{equation*}
\psi (E_i \cdot p) = \psi(p),\quad i=1,\dots, L.
\end{equation*}

\begin{figure}
  \psfrag{f}{$\phi$}
  \centerline{\includegraphics{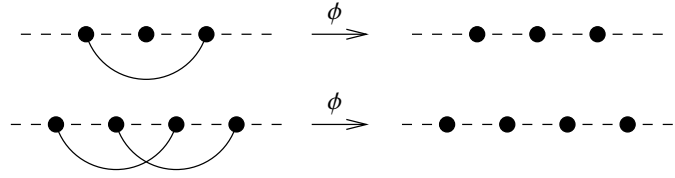}}
  \caption{Illustration of the nontrivial local action of the map $\phi$.}
\label{fig:themapphi}
\end{figure}

To understand the meaning of this invariant let us
define a local map $\phi$ between partial diagrams  which sends
intersected horizontal lines to free points as
depicted in fig.~\ref{fig:themapphi} and acts as identity otherwise.
The local map $\phi$ is applied repeatedly until there are
no more horizontal intersected lines left.
It is not hard to see that $\phi$ extends to a homomorphism of $T_L(1)$
modules
\begin{equation*}
\phi:  Q^l_L(m) \rightarrow D_L(2l+m).
\end{equation*}
In fact, the role of the map $\phi$ is to show that $Q^l_L(m)$ is composed of a direct
sum of isomorphic $D_L(2l+m)$ modules, while that of the map $\psi$ is to
distinguish between these modules.
The set of partial diagrams $p$ in $Q^l_L(m)$ splits into subsets of
constant $\psi(p)$ and each of these subsets is isomorphic to
$D_L(2l+m)$ as a $T_L(1)$ module.

According to what was said before, we get that $n_L(m,j)$ equals to
the number of graphs on $j=m+2l$ vertices and $l$ intersected
edges.
It follows that $n_L(m,j)$ does not actually depend on $L$ and
we drop the index $L$ in the following.
This fact allows, in principle, for an iterative computation of $n(m,j)$ by simply
computing the dimensions of the left and right hand sides of the
decomposition formula
\begin{equation*}
  \Delta_L(m)\simeq \bigoplus_{l=0}^k n(m,m+2l)D_L(m+2l),
\end{equation*}
that is
\begin{equation}
\label{eq:rec1}
  (2k-1)!! C^{2k}_L = \sum_{l=0}^k n(m,m+2l)\left( C^{k-l}_{L-1}- C^{k-l-2}_{L-1}\right)
\end{equation}
successively for $L=0,2,\dots$ or $L=1,3,\dots$.
One can give an explicit expression for $n(m,j)$ with a little
more combinatorial work.

We call a horizontal line an \emph{empty cup} if its ends are adjacent
and simply a \emph{cup} if there are separated by free points.
Observe that all the lines in a partial diagram are intersected
if and only if there is at least one free point in each cup.
Thus, if the partial diagram has $p$ cups with only one free point
inside and a total of $l$ edges
then the remaining $m-p$ free points can be added to the diagram in $C^{2l}_{m-p+2l}$
different ways in such a way that the resulting diagram has only intersected edges.
Moreover, the number of diagrams on $2l$ points with 
$p$ empty cups and a total of $l$ edges is again $n(p,2l-p)$.
This is because the condition of no cups in 
the connection of the remaining $l-p$ edges is similar to the
condition of composing a graph with $p$ free points and $l-p$
intersected edges. 
Putting everything together we get a new recurrence formula
\begin{equation}
\label{eq:rec2}
  n(m,m+2l) = \sum_{p=0}^l C^{2l}_{m-p+2l} n(p,2l-p)
\end{equation}
reducing the problem to the computation of  $n(j):= n(0,2j)$.

Next, we want to find a recurrence relation for $n(j)$ by looking at the
connectivity of the first point in the partial diagrams on $2j$ points
with $j$ intersected edges.
The leftmost vertex in the partial diagram has to be
connected to some other vertex at position $k$.
The connectivity of the $2j-1$ points to the left of the point at
position 1 is equivalent to that in a partial diagram with $j-1$
intersected edges and a free point except for the case where $k=2$.
Therefore we have that
\begin{equation}
\label{eq:rec3}
  n(j) = n(1,2j-1)-n(j-1).
\end{equation}
Now, eq.~\eqref{eq:rec2} yields $n(1,2j-1) = (2j-1)n(j-1)+n(1,2j-3)$.
Using again eq.~\eqref{eq:rec3} for $j-1$ we finally get that
\begin{equation}
  \label{eq:rec4}
  n(j) = (2j-1) n(j-1)+n(j-2).
\end{equation}
The solution of the recurrence eq.~\eqref{eq:rec4} with the initial
conditions  $n(1)=0$ and $n(2)=1$ is
\begin{equation*}
  \label{eq:sol_rec}
  n(j) = \sum_{k=0}^j (-1)^{j-k}\frac{(j+k)!}{2^k (j-k)!k!}
\end{equation*}
and coincides with the absolute value of Bessel polynomials $y_j(x)$
\begin{equation*} 
  y_j(x) = \sum_{k=0}^j\frac{(j+k)!}{(j-k)!k!}\left(
    \frac{x}{2}\right)^{k} 
\end{equation*}
evaluated at $x=-1$.

\section{Conclusion}

Besides the careful definition of the spin model and its sectors, the main
point of this first paper is the algebraic set up necessary to analyze
its symmetries. This is a non trivial task since we are dealing with non
semi-simple algebras,  and that the action of $\OSp(2S+2|2S)$ and
$B_L(2)$ are meshed through a complex structure of indecomposable
representations.
The main results are the decomposition formulas~(\ref{eq:rez1},\ref{eq:rez2})
for $V_{4|2}^{\otimes L}$ viewed as a $\OSp(4|2)$ and a $B_L(2)$
module.
The decomposition in eq.~\eqref{eq:rez1} has been computed in two
essentially different ways: first, by decomposing tensor products
between $\OSp(4|2)$ representations and $V$ without knowing anything
about the Brauer algebra and, second, starting from eq.~\eqref{eq:bf}
with the assumption that the representations of $\OSp(4|2)$ and
$B_L(2)$ on $V^{\otimes L}$ generate the full centralizers of each
other (Schur duality).  The fact that we arrive at the same result
using both methods highly suggests that our assumption about the Schur
duality between $\OSp(4|2)$ and $B_L(2)$ on $V^{\otimes L}$ is correct.

When the question of decomposing $V^{\otimes L}$ is addressed in
sec.~\ref{heart},
the notion of block appears to be a particularly useful
concept for organizing indecomposable representations.\footnote{
Let us note that the blocks appear already in the representation theory
of \emph{simple} Lie algebras if infinite dimensional
representations are allowed. They are precisely the orbits of
the shifted action of the Weyl group on the weight lattice.}
These results will be applied to educated conjectures about   the
conformal field  theory in the next paper. 

Although there are many things left unclear about the representation theory
of $\osp(2S+2|2S),\,S>1$, it is very tempting to
speculate the form of the decomposition of $V^{\otimes L}_{2S+2|2S}$.
Before making the guess, observe that as a $\OSp(4|2)$ module $V^{\otimes L}_{4|2} \simeq T
\oplus P$, where $P$ is a direct sum of projectives organized in blocks,
while $T$ is a direct sum of simples indexed by the
same Young tableau (in the partition notation for dominant weights) as the
irreps of $\OO(2)$. More than that, they appear with the same
multiplicities as their partners in $V^{\otimes
  L}_{2|0}$.\footnote{In is not hard to prove employing the
  methods we used in this paper and the results of \cite{Germoni04}
  for $\osp(3|2)$  that the same phenomenon occurs for $V^{\otimes
    L}_{3|2}$. In this case $T$ is the trivial representation.}
Therefore, $T$ and $V^{\otimes L}_{2|0}$ are similar in all but
the internal structure of their simple summands. The similarity
between the two modules has to be understood in terms of their centralizers, because these
are precisely the objects that do not ``see'' the internal structure
of simples.\footnote{By a corollary of the Schur lemma, if $S$ is a simple
module for the algebra $A$ then $\End_A S  \simeq \mathbb{C}$.}
In conclusion, one should have
$\End_{\OO(2)}V^{\otimes L}_{2|0} \simeq \End_{\osp(4|2)}T$,
which is quite
natural once there is a
Schur duality between $\OSp(4|2)$ and $B_L(2)$ on $V^{\otimes L}_{2S+2|2S}$.
It is tantalizing to speculate  that as a $\OSp(2S+2|2S)$ module
$V^{\otimes L}\simeq T\oplus P$, with $P$ projective and
$\End_{\OSp(2S+2|2S)}T\simeq \End_{\OSp(2S|2S-2)}V^{\otimes
  L}_{2S|2S-2}\simeq B_L(2)/J(S-1)$.
Thus, the problem of the decomposition of $V^{\otimes L}_{2S+2|2S}$ as
a $\OSp(2S+2|2S)$ module is reduced to understanding the projective
representations of the supergroup, i.e. to finding the quiver diagram
for each block. It has been suggested in \cite{Serganova98} that the
quiver diagram of blocks does not depend on $S$ provided the degree of
atypicality $k$ and the action of the outer automorphism $\tau$ are
fixed.\footnote{$\tau$ can act in two ways: either leave invariant all the
 weights in the block or pairwise transform some of them.}
The discussion of sec.~\ref{sec:osp42} suggests that the two types of
quivers for a block of $\osp(2S+2|2S)$ and a fixed $k$ will give rise to
the same quiver for the induced blocks in $\OSp(2S+2|2S)$.

We  also succeeded in computing the multiplicity of Temperley  
Lieb representations in a standard $B_L(2)$-module $\Delta_L(\mu)$.
Finally, we gave a combinatorial description of $B_L(N)$ blocks as
the set of minimal partitions dressed by balanced removable border strips
and have shown that there is a similar description for $\osp(R|2S)$ blocks.

\bigskip

\noindent {\bf Acknowledgments:} This work was supported by the Agence
National Pour la Recherche  under a Programme Blanc 2006 INT-AdS/CFT.
 We would like to thank J. Germoni for fruitful correspondence, and J. Jacobsen and P. Martin
 for useful discussions.

\begin{appendix}

\section{Appendix}

\subsection{$\osp(R|2S)$ Lie superalgebra}\label{sec:superalg}

\begin{figure}
  \psfrag{a1}{$a_1$}
  \psfrag{a2}{$a_{S-1}$}
  \psfrag{a3}{$a_S$}
  \psfrag{a4}{$a_{S+1}$}
  \psfrag{a5}{$a_{S+r-2}$}
  \psfrag{a6}{$a_{S+r-1}$}
  \psfrag{a7}{$a_{S+r}$}
  \centerline{\includegraphics{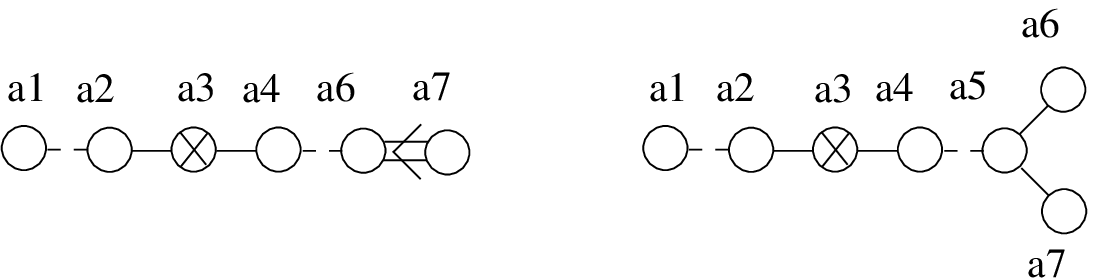}}
  \caption{Distinguished Dynkin diagram for the Lie
    superalgebra $\osp(2r+1|2S)$ on the left and $\osp(2r|2S)$ on
    the right.}
  \label{f:dynkin_diag}
\end{figure}

In this section we recall standard facts about
the $\osp(R|2S)$ Lie superalgebra mainly following the pioneering work of Kac
\cite{Kac77}. For more details on $\osp(R|2S)$ Young supertableaux see
\cite{Sorba85,Hurni87,Jarvis84}.

Let $V$ be a vector space with an additive $\mathbb{Z}_2$ grading
$\dg$, that is $V=V_0\oplus V_1$ and $v\in V_\gamma \Rightarrow \dg(v) =
\gamma$.
Let $\dim V_0 =R$, $\dim V_1=2S$ and $r=[R/2]$.
Choose in $V$ a basis $B=B_0\cup B_1$ with
$B_0=\{ v_i,v^*_i\in V_0,\,(v_{r+1}=v^*_{r+1})\,|\, i=1,\dots,r \}$
and $B_1=\{ u_i,u^*_i\in V_1\,|\, i=1,\dots,2S\}$.
We take the vector $v_{r+1}$ in brackets because it appears for
odd $R$ only. 

The grading of $V$ induces a grading on $\gl(V,\mathbb{C})$, that is
$\gl_0(V,\mathbb{C})$ preserves the degree of $v \in V_\gamma$ and
$\gl_1(V,\mathbb{C})$ changes it.
Define the supertranspose of a matrix $T\in\gl(V,\mathbb{C})$ by
\begin{equation}\label{eq:supertr}
  T=\begin{pmatrix}
    A_{R\times R}& B_{R\times 2S}\\
    C_{2S\times R}& D_{2S\times 2S}
  \end{pmatrix} \quad \Rightarrow \quad
  T^{\strans} = \begin{pmatrix}
    A^{\trans}& C^{\trans}\\
    -B^{\trans} & D^{\trans}
  \end{pmatrix}.
\end{equation}
Let $J$ denote the matrix with the only nonzero
components
\begin{equation}
  \label{eq:j}
  J_{vv^*}=1, \quad J_{v* v}=1,\quad  J_{uu^*}=-1, \quad J_{u*u}=1.
\end{equation}
The Lie superalgebra $\osp(R|2S)$ is realized as a subset of
$\gl(V,\mathbb{C})$ with elements $T$ satisfying 
\begin{equation}\label{eq:salg_def}
  T^{\strans} J + J T = 0.
\end{equation}

In terms of elementary matrices
$(e_{ij})_{kl}=\delta_{ik}\delta_{jl}$ the generators of $\osp(R|2S)$ read
\begin{align}\label{eq:gen1}
  T_{ij}&=e_{ij}-e^{\strans}_{i^*j^*}\\ \label{eq:gen2}
  T_{ij^*}&=e_{ij^*}-(-1)^{\dg(j)}e^{\strans}_{i^*j}\\ \label{eq:gen3}
  T_{i^*j}&=e_{i^*j}-(-1)^{\dg(i)}e^{\strans}_{ij^*},
\end{align}
The generators $h_i=T_{ii}$ span the Cartan subalgebra $\mathcal{H}$.
Denote by $\varepsilon_i$ the basis in  $\mathcal{H}^*$ dual to $h_i$.
It can be easily checked that generators in
eq.~\eqref{eq:gen1} correspond to roots of the type
$\varepsilon_i-\varepsilon_j$,
generators in eq.~\eqref{eq:gen2} correspond to roots of the type
$\varepsilon_i+\varepsilon_j$ and generators in eq.~\eqref{eq:gen3}
correspond to roots of the type $-\varepsilon_i-\varepsilon_j$.
The bilinear invariant form $-\frac{1}{2}\str (h_i h_j)$ induces a
scalar product on $\mathcal{H}^*$.

The standard basis is recovered by putting $\epsilon_i=\varepsilon_i$ for
$i=1,\dots,r$ and $\delta_i=\varepsilon_{r+i}$ for $i=1,\dots,S$.
Elementary weights $\delta_i,\epsilon_j$ are orthogonal in
$\mathcal{H}^*$ and $\delta_i^2=-\epsilon_i^2=1$.
The first $r+S-1$ simple roots are chosen to be
$\alpha_{i}=\delta_i-\delta_{i+1}$, $\alpha_S=\delta_n-\epsilon_1$,
$\alpha_{S+j}=\epsilon_j-\epsilon_{j+1}$ for $i=1,\dots,S$ and
$j=1,\dots,r-1$.
The last simple root is $\alpha_{r+S}=\epsilon_r$ for odd $R$ and
$\alpha_{r+S}=\epsilon_{r-1}+\epsilon_r$.
The roots $\pm \delta_i\pm\epsilon_j$ are called odd and the rest --- even.

The component of a weight $\Lambda$ along the hidden simple
$\ssp(2S)$ root $2\delta_S$ is
\begin{align}
  \label{eq:longest_sp_root}
\textnormal{$R$ odd}:\quad b &= a_S - a_{S+1} - \dots - a_{S+r-1} -
a_{S+r}/2\\
\textnormal{$R$ even}:\quad b &= a_S - a_{S+1} - \dots - a_{S+r-2} - (
a_{S+r-1}+  a_{S+r})/2.
\end{align}
According to \cite{Kac77}, an $\osp(R|2S)$ highest weight is dominant iff
it has integer Dinkyn labels $a_{i\neq S}$ and integer $b$ satisfying
the following consistency conditions
\begin{align} \label{eq:consistency}
  \textnormal{$R$ odd}:\quad b &\leq r-1 \Rightarrow
  a_{S+b+1}=\cdots=a_{S+r}=0\\ \notag
  \textnormal{$R$ even}:\quad b &\leq r-2 \Rightarrow
  a_{S+b+1}=\cdots=a_{S+r}=0, \quad b = r-1 \Rightarrow a_{S+r-1}=a_{S+r}=0 \notag.
\end{align}
  All irreducible finite dimensional representations are indexed by
  dominant weights $\Lambda$. 
Given a dominant weight $\Lambda=\sum \rho_i \delta_i +\sum
 \sigma_j\epsilon_j$ in the standard basis, the first $r+S-1$ Dynkin labels are
$a_i = \rho_i - \rho_{i+1}$ for $i=1,\dots, S$,
$a_{S+i}=\sigma_i-\sigma_{i+1}$ for $i=1,\dots,r-1$.
The last Dynkin label is
$a_{S+r}=2\sigma_r$ for $R$ odd and $ a_{S+r} =
\sigma_{r-1}+\sigma_{r}$ for $R$ even. 
From eq.~\eqref{eq:longest_sp_root} we also get $b=\rho_S$.

\begin{figure}
      \psfrag{q}{$\longrightarrow$}
      \psfrag{q0}{$\times$}
      \psfrag{q1}{$\rho_1$}
      \psfrag{q2}{$\rho_2$}
      \psfrag{q3}{$\rho_{r-1}$}
      \psfrag{q4}{$\rho_{r}$}
      \psfrag{q5}{$\sigma_1$}
      \psfrag{q6}{$\sigma_2$}
      \psfrag{q7}{$\sigma_{S-1}$}
      \psfrag{q8}{$|\sigma_{S}|$}
      \centerline{\includegraphics{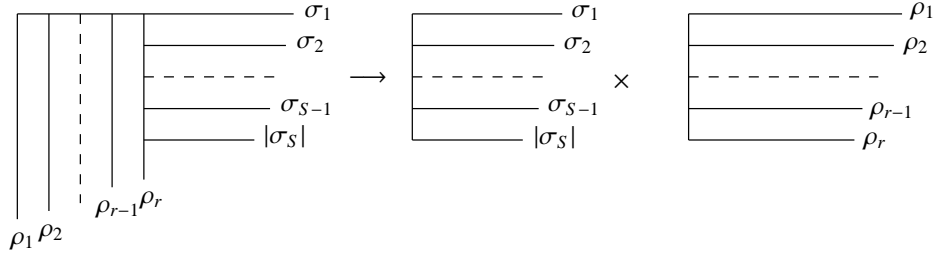}}
      \caption{The $\so(R)\times \ssp(2S)$ representation
	to which belongs the highest weight state of a $\osp(R|2S)$
	representation.}
      \label{f:hook_shape}
\end{figure}

The set of numbers
$\rho_i,\sigma_j$ define a
partition, shown in fig.~\ref{f:hook_shape}, provided that consistency
conditions \eqref{eq:consistency} plus some 
additional constraints depending on $R$ are satisfied.
These additional constraints require $a_{S+r-1}< a_{S+r}$ and
$a_{S+r-1}+a_{S+r}$ to be even 
if $R$ is even, and $a_{S+r}$ to be
to be even if $R$ is odd.
The last two conditions define tensorial weights.

Partitions $\lambda$ such that
$\lambda_{r+1}\leq S$ are called hook shape.
Let $\tau$ denote the outer automorphism induced by the symmetry of
the $\osp(2r|2S)$ Dynkin diagram under the exchange of the last two roots
in fig.~\ref{f:dynkin_diag}.
This automorphism is extremely important in understanding the
difference between the representation theory of the supergroup
$\OSp(R|2S)$ and its Lie superalgebra.
Note that $\tau$ can be explicitly realized through the
discrete transformation $\rho$ exchanging the last two basis vectors
in $B_0$. Indeed, $\rho(\epsilon_j) = \epsilon_j$ for $j=1,\dots,r-1$
and $\rho(\epsilon_r) = -\epsilon_r$ because $\epsilon_j$ are the duals
of $e_{jj}-e_{j^*j^*}$.
Therefore $\rho(\alpha_{S+r})=\rho(\epsilon_{r-1}+\epsilon_r) =
\epsilon_{r-1}-\epsilon_r=\alpha_{S+r-1}$.

In the case of $R$ odd, there is a bijective correspondence between
hook shape partitions
$\lambda$ and dominant weights $\Lambda$.
The same holds for $R$ even, except for $\lambda$ with $\sigma_r>0$
when $\lambda$ represents both $\Lambda$ and $\tau \cdot \Lambda$.

%  Let $V(\Lambda)$ denote a Kac module.
%     Under the action of the even part the
%     $\osp(R|2S)$ module $V(\Lambda)$ decomposes into
%     $\so(2m)\times \ssp(2n)$ multiplets. Even generators of the
%     superalgebra act inside a multiplet and odd generators connect
%     multiplets. The maximum number of multiplets is the number $2mn$ of
%     odd roots. When smaller, some multiplets either are projected to zero
%     by the action of generators or they generate an invariant subspace of
%     $V(\Lambda)$. This happens iff there are $1\leq i\leq m$ and $1\leq
%     j\leq n$ such that at least one of the conditions 

If there is a pair $(i,j)$ such that at least
one of the conditions below are satisfied
\begin{align}\label{e:1_atyp}
  &\rho_j +\sigma_i + S + 1-i-j=0 \\ \label{e:2_atyp}
  &\rho_j -\sigma_i + S - R +1-j+i=0,
\end{align}
the weight $\lambda$ is called \emph{atypical}.\footnote{For $b\leq r-1$ the
  highest weight $\Lambda$ is always atypical.}
See \cite{Kac78} for
the origin of these conditions and note ref.~\cite{Jarvis84}, where
these have been presented in the form (\ref{e:1_atyp},\ref{e:2_atyp}).
If none of these conditions is satisfied, the weight is called
\emph{typical} and, according to \cite{Kac78},
the associated Kac module $\Bar{V}(\Lambda)$ (which is a finite
dimensional quotient of the
corresponding highest weight module) is
irreducible, its (super)character is  given by the Weyl-Kac formula
\cite{Kac77a} and, in particular, its superdimension is zero.
%
% To  decompose $\Bar{V}(\Lambda)$ as a  $\so(R)\times \ssp(2S)$ module
% one can use the Poincar\'{e}-Birkhoff-Witt theorem.
% For $\Lambda$ typical, $\Bar{V}(\Lambda)$ decomposes (at most)
%  into $2^{RS}$ irreps of $\so(R)\times \ssp(2S)$.

\subsection{$\OSp(R|2S)$ supergroup}
\label{sec:supergr}

Let $\Gamma = \Gamma_0 \oplus \Gamma_1$ be a Grassman algebra. 
 The supergroup $\OSp(R|2S)$ may be
 realized as a subset of even supermatrices
    \begin{equation*}\label{eq:supermat_form}
      M=\begin{pmatrix}
	A_{R\times R}& B_{R\times 2S}\\
	C_{2S\times R}& D_{2S\times 2S}
	\end{pmatrix},
    \end{equation*}
    with entries in $A$ and $D$ belonging to $\Gamma_0$, and
    entries in $B$ and $C$ belonging to $\Gamma_1$, which satisfies
    \begin{equation}
      \label{eq:sgr_def}
      M^{\strans} J M = J.
    \end{equation}
Equivalently, $\OSp(R|2S)$ can be seen as the set of linear
transformations leaving invariant the graded symmetric form
\begin{equation}
  \label{eq:inv_form}
  \eta_1 . \eta_2 := \eta^{\trans}_1 J \eta_2 = \sum_{i=1}^r
   b^{i^*}_1
  b^i_2 +b^i_1 b^{i^*}_2  + \left(
  b^2_{r+1}\right)+\sum_{j=1}^S f^{j^*}_1f^j_2-f^j_1f^{j^*}_2,
\end{equation}
where $\eta_\alpha$  are arbitrary points in a superspace
parametrized by coordinates
$b^i_\alpha,b^{i^*}_\alpha \in \Gamma_0$ and
$f^j_\alpha, f^{j^*}_\alpha \in  \Gamma_1$ and $\alpha =1,2$.

Representing $M=I+\sum _a \alpha_a T_a$ with infinitesimal $\alpha_a
\in \Gamma_0,\Gamma_1$ and expanding eq.~\eqref{eq:sgr_def} one gets
the definition \eqref{eq:salg_def} of the superalgebra $\osp(R|2S)$.
Thus, the subgroup of $\OSp(R|2S)$ connected to identity is an
exponential of $\osp(R|2S)$. The representation theory of both is
the same as long as we restrict to tensor representations which are
the only ones appearing in the tensor space $V^{\otimes L}$.
 
From the definition~\eqref{eq:sgr_def}
any matrix $M\in \OSp(R|2S)$ has superdeterminant $\sdet M =\pm 1$.
The supergroup has two
disconnected parts $\OSp^\pm(R|2S)$, which  correspond to the value of the
superdeterminant of its elements, that is $\OSp(R|2S)/\OSp^+(R|2S)=\mathbb{Z}_2$.

To see this, one can repeat the same reasoning typical of $\OO(N)$ groups.
Elementary transformations susceptible to change the sign of the
superdeterminant belong to the discrete symmetry group $W$ of  
the $\OSp(R|2S)$ invariant form \eqref{eq:inv_form}.
The generators of $W$
are read out from eq.~\eqref{eq:inv_form} to be ``reflections''
$\rho_i:(b_i,b^*_i)  \mapsto (b^*_i,b_i)$ and 
$\rho'_j:(f_j,f^*_j) \mapsto (- f^*_j,f_j)$, and permutations
$\pi_i:(b_i,b^*_i) \leftrightarrow (b_{i+1},b^*_{i+1})$ and  $\pi'_j:
(f_i,f^*_i) \leftrightarrow (f_{i+1},f^*_{i+1})$.
For odd $R$ there is also the reflection $\rho_{r+1}:b_r\mapsto -b_r$.
The subgroup $W$ is in fact the Weyl group of the root system of $\so(R)\times \ssp(2S)$.
Denote by $W^\pm$ the set of elements of $W$ embedded in $\OSp^\pm(R|2S)$.
It is easy to see that all elements of $W^-$ are conjugate in
$W^+$ to a single reflection $\rho$, which one can take $\rho_r$ if $R$
is even and $\rho_{r+1}$ if $R$ is odd.
Therefore, we see that indeed $W/W^+=\mathbb{Z}_2$.

Let $v_{\Lambda'} \in g(\Lambda)$ be a vector of weight $\Lambda'\leq \Lambda$.
Then, as seen in sec.~\ref{sec:superalg}, there is an action of $\rho$
on $g(\Lambda)$ provided by $\rho \cdot v_{\Lambda'} = v_{\tau \cdot
  \Lambda'}$.
In the case of $\osp(4|2)$ the outer automorphism $\tau$ exchanges
$\epsilon_2$ with $\epsilon_3$.
The representations induced from $\osp(R|2S)$ to $\OSp(R|2S)$ are
of the form
\begin{equation}\label{eq:coh}
\OSp(R|2S) \otimes_{\OSp^+(R|2S)} g(\Lambda)\simeq \mathbb{Z}_2
\otimes_\rho g(\Lambda).
\end{equation}
There are two possible cases now:
i) either $\rho \cdot g(\Lambda)= g(\Lambda) \Leftrightarrow \tau\cdot
\Lambda = \Lambda$  and then obviously  $\mathbb{Z}_2 \otimes_\rho
g(\Lambda)= 1\otimes_\rho g(\Lambda) \bigoplus \varepsilon \otimes_\rho g(\Lambda)$
with $\rho\cdot 1 =1$ and $\rho\cdot \varepsilon=-\varepsilon$ or ii)
$\rho \cdot g(\Lambda)\neq g(\Lambda)\Leftrightarrow \tau\cdot \Lambda\neq \Lambda$ and the induced
module in eq.~\eqref{eq:coh} is irreducible.

Two representations $R(\rho), R^*(\rho)=-R(\rho)$ of $\mathbb{Z}_2$
are called \emph{associate}.
The modules $G(1\times \lambda):= 1\otimes_\rho g(\Lambda)$ and
$G(\varepsilon\times \lambda):=\varepsilon \otimes_\rho g(\Lambda)$
are also called associate. 
In contrast, $G(\tau\times \lambda):=\mathbb{Z}_2\otimes_\rho
g(\Lambda)\simeq \mathbb{Z}_2\otimes_\rho g(\tau\cdot \Lambda)$ is
isomorphic to its associate because there is an
equivalence transformation between $R(\rho)$ and $R^*(\rho)$ through
the change of sing of basis vectors in the subspace $\rho \otimes
g(\Lambda)$.
Therefore $G(\tau\times \lambda)$ is called selfassociate.

A direct implication following from the definitions of (self) associate
modules is $\sch_{\mu}(D) = \sdet D\sch_{\mu^*}(D)$, where $\mu,\mu^*$
are (self)associate weights of $\OSp(R|2S)$.
For a  selfassociate weight $\mu$ this equality implies $\sch_\mu(D) =
0$ if $\sdet D = -1$.

Note that the centralizer of $B_L(N)$ on $V^{\otimes L}$
is the direct product algebra $\mathbb{Z}_2 \times \osp(R|2S)$ rather
then $\osp(R|2S)$.
This algebra has the same tensor irreducible representations as the
supergroup $\OSp(R|2S)$.

\subsection{$\osp(4|2)$ Lie superalgebra and $\OSp(4|2)$ supergroup}
\label{sec:osp42}

This is a compact resum\'{e} of the results presented in
\cite{Germoni04} plus some additional remarks on the representation
theory of $\OSp(4|2)$.

The superalgebra $\osp(4|2)$ has minor differences with respect to
the general context of $\osp(R|2S)$ superalgebras,
because of the isomorphism $\so(4)\simeq \ssl(2)\times \ssl(2)$.
The even part of the superalgebra is $\so(4)\times\ssp(2)\simeq
\ssl(2)\times\ssl(2)\times\ssl(2)$.
The odd  part is a representation of the even part of dimension
$2\times 2\times 2$.

The standard basis vectors
$\{\epsilon_1,\epsilon_2,\epsilon_3\}$  of $\mathcal{H}^*$ are normalized as $\epsilon_1^2 =
-1,\,\epsilon_2^2=\epsilon_3^2=1/2 $.
The even and the odd positive root systems are $\Delta^+_{0}=\{2 \epsilon_1,2
\epsilon_2,2 \epsilon_3\}$ and $\Delta^+_1 = \{\epsilon_1 \pm
\epsilon_2\pm \epsilon_3\}$.
The simple roots are traditionally chosen as $\alpha_1 = \epsilon_1 -
\epsilon_2 - \epsilon_3$, $\alpha_2=2\epsilon_2$,
$\alpha_3=2\epsilon_3$.
The hidden root will then be
$2\epsilon_1 = \alpha_1 + \alpha_2 + \alpha_3$.

Consistency conditions
\eqref{eq:consistency} for a dominant weight $\Lambda= b\epsilon_1 +
a_2 \epsilon_2 + a_3 \epsilon_3$,
require $b=0 \Rightarrow a_2=a_3=0$
and $b=1 \Rightarrow a_2=a_3$.
%
% The half sum of positive even roots is
% $\rho_0=\epsilon_1+\epsilon_2+\epsilon_3$, of positive odd roots is
% $\rho_1= 2\epsilon_1$ and
% $\rho\equiv\rho_0-\rho_1=\epsilon_2+\epsilon_3-\epsilon_1$.
%
We associate to $\Lambda$ a hook shape partition $\lambda$
with symplectic part $\rho_1=b$ and orthogonal part $\sigma_1=(a_2+a_3)/2,
\sigma_2= |a_2-a_3|/2$.
To make the correspondence $\Lambda
\rightarrow \lambda$ bijective we mark the partition $\lambda$
by $\sgn(\sigma_1-\sigma_2)$ when  $\lambda_2>1$.\footnote{ Any $\so(4)=\ssl(2)\oplus\ssl(2)$ irreps can be written as a couple
$(j_1,j_2)$ of $\ssl(2)$
  irreps.
  The sign attached to $\lambda$ distinguishes between
  $(j_1, j_2)$ and $(j_2, j_1)$ when $ j_1\neq j_2$.}

% \begin{figure}
%   \psfrag{b}{$b$}
%   \psfrag{c}{$(a_2+a_3)/2$}
%   \psfrag{d}{$|a_2 - a_3|/2$}
%   \psfrag{e}{$\sgn (a_2 - a_3)$}
%   \centerline{\includegraphics{fig/21hook.eps}}
%   \caption{The Young tableau $[\lambda]$ marked with a sign 
%     associated to the weight $\Lambda = (b,a_2,a_3)$.}
%   \label{f:21hook}
% \end{figure}

Atypicality conditions~(\ref{e:1_atyp},\ref{e:2_atyp}) take
the form
\begin{align*}
  \rho_1+\sigma_1&=0,\quad \rho_1+\sigma_2-1=0\\
  \rho_1-\sigma_1-2&=0,\quad \rho_1-\sigma_2-1=0.
\end{align*}
The solutions can be parametrized by two integers $k$ and $l$.
For $k=0$ these are $\lambda_{0,l}=l1^l, l\geq 0$, while for
$k>0$, $\lambda_{k,1}=k, \lambda_{k,l}=kl1^{l-2}, 2\leq l\leq k$
and $\lambda_{k,l}=l(k+1)1^{l-1}, k+1\leq l$.

% 
% The structure of $V_\Lambda$ can be read from fig.~\ref{f:diag_ind}.
% The dimensions of $V(\Lambda)$ can be computed from
% fig.~\ref{f:diag_ind} and the dimension for simples --- $\dim
% V(\lambda_2) = 1+17+110=128$, $\dim V(\lambda_l)=16l^3$ and $\sdim
% V(\lambda_l)=0$, $l\geq 2$.
% 

% 
% Indecomposables have
% dimension $\dim V(\lambda^{\pm}_{k,l})=16l|k^2-l^2|$ and superdimension $\sdim
% V(\lambda^{\pm}_{k,l})=0$, $l>0$.
% 
\begin{figure}
  \psfrag{0}{$\lambda_{0,0}$}
  \psfrag{1}{$\lambda_{0,1}$}
  \psfrag{2}{$\lambda_{0,2}$}
  \psfrag{3}{$\lambda_{0,4}$}
  \psfrag{4}{$\lambda_{0,5}$}
  \psfrag{k0}{$\lambda_{k,1}$}
  \psfrag{k1}{$\lambda_{k,2}^+$}
  \psfrag{k2}{$\lambda_{k,2}^-$}
  \psfrag{k3}{$\lambda_{k,3}^+$}
  \psfrag{k4}{$\lambda_{k,3}^-$}
  \psfrag{k5}{$\lambda_{k,4}^+$}
  \psfrag{k6}{$\lambda_{k,4}^-$}
  \centerline{\includegraphics{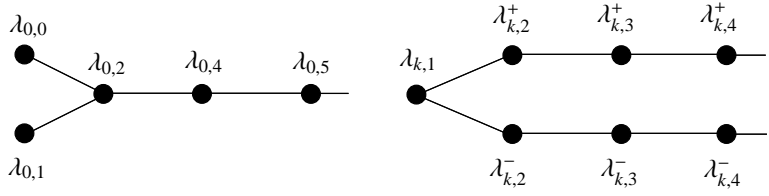}}
  \caption{The quiver diagrams of type $D_\infty$ and $A^\infty_\infty$ for the blocks of $\osp(4|2)$.}
  \label{f:diag_ind}
\end{figure}

Denote by  $g(\lambda)$ the  $\osp(4|2)$ simple modules.
For $\lambda$ typical  $g(\lambda)\simeq \Bar{V}(\lambda)$ and $\dim
V(\lambda)$ can be computed by decomposing $\Bar{V}(\lambda)$ into (at
most 16) representations of $\ssl(2)^{\times 3}$
\begin{equation*}
  \dim g(\lambda)=16(b-1)(a_2+1)(a_3+1).
\end{equation*}
The dimensions of $g_{k,l}:= g(\lambda_{k,l})$ can be computed with
the help of character formulas given in \cite{Germoni04}.

For $k=0$ we get $\dim g_{0,0}=\sdim g_{0,0}=1$, $\dim g_{0,1}=17,\sdim
g_{0,1}=1$ and $\dim
g_{0,l}=D_l^3-3D_l, \sdim g_{0,l}=2, l\geq 2$, where $D_j = 2j+1$.

For $k>0$ we get $\dim g_{k,1}=4k^2+2, \dim
g^\pm_{k,l} =  D_k D_{k-1} D_{l-1} -  D_{l-1}^2 D_{l-2}-
2D_{l-1} D_{l-2}, 2\leq l\leq k$ and $\dim g^\pm_{k,l}= D_l^2 D_{l-1}
+ 2D_l D_{l-1} -
D_k D_{k-1} D_l, l\geq k+1$, and $\sdim g_{k,0}=2, \sdim g^\pm_{k,l}=2, l\geq 1$.

The set of weights $\lambda_{k,l}$ with $k$ fixed belong to the same block of
$\osp(4|2)$.
To (at least partially) see this one has to check that the second
order Casimir invariant takes the same value $k^2$ on the whole block $k$.\footnote{The
  second order Casimir is a central element of the enveloping
  superalgebra. The eigenvalues of central elements on $g(\lambda)$ define the
\emph{central character} of $g(\lambda)$.
If two weights $\lambda,\lambda'$ are in the same block then $g(\lambda),g(\lambda')$
have the same central characters.
This is a consequence of the extension of the Schur lemma (in the form known to
physicists) to indecomposable representations.}
The actual construction of the set of indecomposable modules
providing the equivalence relation of sec.~\ref{sec:tmb} between the weights of a block is
done in \cite{Germoni04}.

The quiver diagram representing the structure of 
$\osp(4|2)$ projective modules in a block is represented in
fig.~\ref{f:diag_ind}.
The projective covers $\mathcal{P}g_{k,l}$ of the modules $g_{k,l}$
in the block $k=0$ have the submodule structure 
\begin{align}\notag
  g_{0,0}&  \quad& g_{0,1}& \quad& g_{0,2}&  \quad& g_{0,l}& \quad& &\\
  g_{0,2}&  \quad& g_{0,2}& \quad& g_{0,0}\;g_{0,1}&\;g_{0,3}
  \quad& g_{0,l-1}\;g&_{0,l+1} \quad& l\geq 3,& \label{eq:projj} \\
  g_{0,0}&  \quad& g_{0,1}& \quad& g_{0,2}&  \quad& g_{0,l}&
  \quad& & \notag
\end{align}
while in the block $k>0$ their submodule structure is
\begin{align}\notag
  g^-_{k,l}&  \quad& g_{k,1}& \quad& g^+_{k,l}&  \quad& \\
  g^-_{k,l-1}\;g&^-_{k,l+1} \quad&  g^-_{k,2}\;g&^+_{k,2} \quad&
  g^+_{k,l-1}\;g&^+_{k,l+1}, \quad& l\geq 2. \label{eq:projjj} \\
  g^-_{k,l}&  \quad& g_{k,1}& \quad& g^+_{k,l}&  \quad& \notag
\end{align}

The dimensions of projective modules in the block $k=0$ are
$\dim\mathcal{P}g_{0,0}=112, \dim
\mathcal{P}g_{0,l}=16(2l+1)(1+l+l^2),l\geq 1$,
while in the block $k>0$ there are $\dim \mathcal{P}g_{k,1}=32(k^2-1),
\dim \mathcal{P}g_{k,l}=16(2l-1)(k^2-1+l-l^2), l\leq k-1,
\dim\mathcal{P}g_{k,k}=32(1+2k^2),
\dim\mathcal{P}g_{k,l}=16(2l+1)(1-k^2+l+l^2), l\geq k+1$.
The superdimension of projective (including typical) modules vanishes.

Let us apply the general discussion of sec.~\ref{sec:supergr} to the
supergroup $\OSp(4|2)$. The outer automorphism $\tau$
acts on $\mathcal{H}^*$ by exchanging $\epsilon_2$ with $\epsilon_3$.
Consequently, $\rho \cdot g_{k,l} = g_{k,l}$ for $k=0$ or $l=0$ and $\rho
\cdot g^\pm_{k,l}=g^\mp_{k,l}$ otherwise.

\begin{figure}
  \centerline{\includegraphics{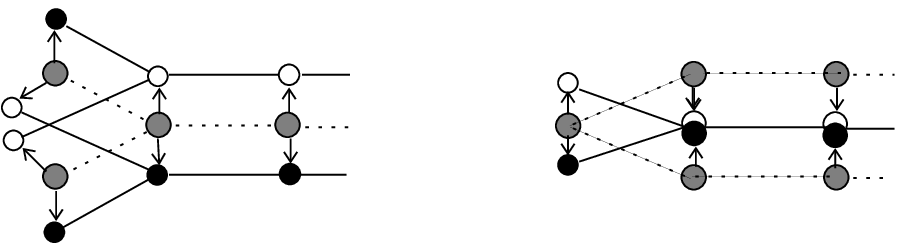}}
  \caption{Schematic picture showing how the induction
    procedure (vertical arrows) sends the quiver
    diagrams of type $D_\infty$ and $A^\infty_\infty$  (grey dots and
    dotted lines) for $\osp(4|2)$ blocks, represented in
    fig.~\ref{f:diag_ind}, into quiver diagrams of type
     $D_\infty$ for $\mathbb{Z}_2\times \osp(4|2)$ blocks.
    White and black dots represent weights of the type $1\times
    \lambda$ and $\varepsilon\times \lambda$ respectively. Double
    circles represent selfassociate weights  $\tau\times \lambda$.}
  \label{fig:cool}
\end{figure}

We claim that the quiver diagram of type $D_\infty$ for the block $k=0$ of $\osp(4|2)$ will
give rise to two quiver diagrams of type $D_\infty$, as shown in
fig.~\ref{fig:cool}, and,
consequently, to two associate blocks for the algebra
$\mathbb{Z}_2\times \osp (4|2)$, which we call $k=0,0^*$.
As we shall see bellow, the weights in the block $k=0$ are $1\times \lambda_{0,0},
\varepsilon\times\lambda_{0,l},l\geq 1$, while the weights in the block
$k=0^*$ are $\varepsilon\times
\lambda_{0,0},1\times\lambda_{0,l},l\geq 1$.

We also claim that the quiver diagram of type
$A^\infty_\infty$ for the block $k\neq 0$ of $\osp(4|2)$ will give rise to a
single quiver diagram of type $D_\infty$, as shown in
fig.~\ref{fig:cool}, and a selfassociate block for
the algebra $\mathbb{Z}_2\times \osp(4|2)$, which we label also by
$k$. As we shall sea bellow, the weights in the block $k\neq 0,0^*$ are $1\times
\lambda_{k,0},\varepsilon\times
\lambda_{k,0},\tau\times\lambda_{k,l},l\geq 1$.

Our claim follows from the analysis of shift operators
$A^{\alpha\beta\gamma},\,\alpha,\beta,\gamma=\pm$
introduced in \cite{Jeugt85}.
These operators are very practical for decomposing  Kac modules
$\bar{V}(\lambda)$ into $\ssl(2)^{\times 3}$ irreps.\footnote{The fact
  choose Kac modules in order to understand the transformation properties under
  $\rho$ of arrows in the quiver diagram of a block is irrelevant for the following.
 One can take instead of $\bar{V}(\lambda)$ the standard modules
 $\mathcal{L}_0(\lambda)$ as well, which are defined 
 by cohomological induction in \cite{Germoni04}.}
To be more
specific, let  $v_\mu\in \bar{V}(\lambda)$ be a vector of weight $\mu$
maximal for the algebra $\ssl(2)^{\times 3}$.
Then,  $A^{-\beta\gamma}v_\mu\neq 0$ is again a maximal vector of weight
$\mu -\epsilon_1 +\beta\epsilon_2+\gamma\epsilon_3$ for
$\ssl(2)^{\times 3}$.
To identify $\osp(4|2)$ irreducible components in $\bar{V}(\lambda)$
one has to search for $\ssl(2)^{\times 3}$ maximal vectors with dominant
weights $\mu$ in the same block as $\lambda$.

Consider first the block $k=0$ of $\osp(4|2)$.
Then $\Bar{V}(\lambda_{0,l}),l\geq2$ has a $\ssl(2)^{\times 3}$
maximal vector $A^{---}v_{\lambda_{0,l}}$ and one can check that all
positive odd generators annihilate it. Therefore
$A^{---}v_{\lambda_{0,l}}$ is a maximal vector for $\osp(4|2)$ and
$\bar{V}(\lambda_{0,l})$ contains at least $g_{0,l}$ and $g_{0,l-1}$.
In fact, these are the only two irreducible factors of
$\bar{V}(\lambda_{0,l}),l\geq 3$ because, using the Weyl-Kac formula
for characters and the results of sec.~\ref{sec:osp42},
one can check that $\dim\bar{V}(\lambda_{0,l})=\dim g_{0,l}+\dim g_{0,l-1}$. In the
case $l=2$ one has that $\dim \bar{V}(\lambda_{0,2})-\dim
g_{0,2}-\dim g_{0,1} =1$ and, therefore, $\bar{V}(\lambda_{0,2})$
contains also the trivial representation.

In order to see how the four weights $1\times \lambda_{0,l},1\times
\lambda_{0,l-1}, \varepsilon\times \lambda_{0,l},\varepsilon\times
\lambda_{0,l-1}$ split into two different blocks  of $\mathbb{Z}_2\times
\osp(4|2)$ one has to check out how $A^{---}$ transforms under the action of $\rho$.
From the explicit expression of shift operators in \cite{Jeugt85} it
follows that $\rho A^{\alpha\beta\gamma}\rho =
A^{\alpha\gamma\beta}$
and, consequently,  $1\times \lambda_{0,l},1\times 
\lambda_{0,l-1}$  are in the same block and $\varepsilon\times
\lambda_{0,l},\varepsilon\times \lambda_{0,l-1}$ are in an other
same block of $\mathbb{Z}_2\times \osp(4|2)$.

The module $\Bar{V}(\lambda_{0,2})$ has a maximal vector
$A^{---}A^{-+-}A^{--+}v_{\lambda_{0,2}}$, of weight zero, corresponding
to the trivial representation $g_{0,0}$. With the help of relations in
appendix \cite{Jeugt85} for the shift operator products of type
$(1,0,0)$, one can show that $\rho
A^{---}A^{-+-}A^{--+}v_{\lambda_{0,2}} =
A^{--+}A^{-+-}A^{---}v_{\lambda_{0,2}} =
-A^{-+-}A^{--+}A^{---}v_{\lambda_{0,2}}$ and, therefore, $g_{0,0}$
belongs to the block $k=0$ of  $\mathbb{Z}_2\times \osp(4|2)$ as claimed.

Consider now the block $k>0$ of $\osp(4|2)$.
Then $\bar{V}(\lambda^\pm_{k,l})$ will have a single $\osp(4|2)$
maximal vector (besides $v_{\lambda^\pm_{k,l}}$) corresponding to the
irrep $g^\pm_{k,l-1}$ given by $A^{---}v_{\lambda^\pm_{k,l}}$ if
$l>k+1$,  $A^{---}A^{-\mp\pm}v_{\lambda^\pm_{k,k+1}}$ if $l=k+1$
 and $A^{-\mp\pm}v_{\lambda^\pm_{k,l}}$ if $2 \leq l\leq k$.
The induced module $\mathbb{Z}_2\otimes_\rho
\bar{V}(\lambda^+_{k,l}),\,l\geq 2$ will be the sum of 
$\bar{V}(\lambda^\pm_{k,l})$ glued together by the action of $\rho$.
Finally, the induced module
$\mathbb{Z}_2\otimes_\rho\bar{V}(\lambda^+_{k,2})$ has two irreducible
components $1\otimes g_{k,0}$ and  $\varepsilon\otimes g_{k,0}$ with $\mathbb{Z}_2\times \osp(4|2)$
maximal vectors $(1\pm\rho)\otimes A^{--+}v_{\lambda^+_{k,2}}$.

\section{Blocks, minimality and atypicality}
\label{sec:tmb}

In this section we  explain carefully the notion of block
appearing in the representation theory of nonsemisimple algebras.
We also look in details at the similarity between the blocks of $\osp(R|2S)$ and $B_L(N)$.
%
%One should be aware from the start that there are not exactly the
%same, because we expect $\mathbb{Z}_2\times \osp(R|2S)$ to be  the %full centralizer of
%$B_L(N)$ on $V^{\otimes L}$.

In the representation theory of non semisimple algebras the block is
an essential notion.
The \emph{blocks} are conjugacy classes of irreps with respect to the
equivalence relation $\equiv$ defined as follows.
Let $\mathcal{I}$ be the category of indecomposable modules of the algebra.
Write $S_1\equiv S_2$ if there is an indecomposable module in
$\mathcal{I}$ with simple summands $S_1,S_2$. Extend the relation
$\equiv$ by transitivity in order to get an equivalence.
In a semisimple algebra the notion of block is irrelevant because
indecomposable representation are irreducible and the congruence $\equiv$ becomes an
equality.

A relevant example is the Temperley Lieb algebra, with fugacity for
loops $N$ in its adjoint/diagrammatic representation.
For generic values of $N$ the algebra  is semisimple and, thus, has only
completely reducible representation.
Restricting to subsets of planar diagrams, with the number of vertical lines
fixed to $m$, and treating all the other diagrams as zero,
we get all irreps, which are parametrized by $m$.
However, at special points $N=2\cos \pi r'/r''$ with coprime integers
$r',r''$, the algebra becomes
nonsemisimple, irreps  labeled by $m$ become reducible and $m$
becomes a label for a whole block of the algebra, see \cite{Martin}.

The irreducible components $B_L(\lambda)$ of indecomposable modules
 $\Delta_L(\mu),$ when the Brauer algebra $B_L(N)$ is nonsemisimple, where
first studied by mathematicians  Hanlon \emph{et al}
in \cite{Hanlon99}.
Recently Martin \emph{et al}  gave a complete
description for the blocks of the Brauer algebra in  \cite{Martin06}.

We  introduce the same notation as in \cite{Martin06} to formulate
their block result for $B_L(N)$.
If the box $\epsilon$ is in the row $i$ and column $j$ of the Young
tableau of a partition $\mu$, then
its content is $c(\epsilon)=j-i$.
Two boxes $\epsilon,\epsilon'\in \lambda$ are called balanced if
$c(\epsilon)+c(\epsilon')=1-N$.
For two partitions $\mu\subset \lambda$,
the skew partition $\lambda/\mu$ is called
balanced if it is composed of balanced pairs of boxes.

\begin{figure}
   \psfrag{a}{$a$}
   \psfrag{b}{$b$}
     \centerline{\includegraphics{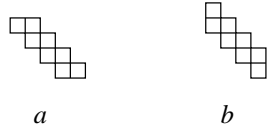}}
      \caption{Two possible configurations of boxes with content
        $1-N/2,-N/2$.}
      \label{fig:2conf}
    \end{figure}

The necessary condition for $\Delta_L(\mu)$ to contain  $B_L(\lambda)$ is:
i) $\mu\subset \lambda$ and $\lambda/\mu$ is balanced;
ii)
If $N$ is even and the boxes of content $1-N/2,-N/2$ in
  $\lambda/\mu$ are configured as shown in case $a$ fig.~\ref{fig:2conf}, then
  the number of columns in this configuration is even.

The given necessary criterion has the structure of a partial ordering.
If $\mu\subset\lambda$ satisfy i)
and ii) we write $\mu\preceq \lambda$.
The splitting of the set of weights $X_L$ into posets
with respect to $\preceq$ gives the blocks of $B_L(N)$.
As shown in \cite{Martin06}, there is a unique minimal partition in a block, which can serve
as a label.

A sufficient criterion for the module
$\Delta_L(\mu)$ to contain  $B_L(\lambda)$ was derived in
\cite{Martin06} and  requires $\lambda$ to be the least 
weight $\lambda\succeq\mu$. 

We want to give a combinatorial description of the
weights in a block.
Consider the Young tableau of a partition $\lambda$ in the block of
the minimal partition $\mu$.
Let $\epsilon_1$ ($\epsilon_1'$) be the box with the highest (lowest) content
in the skew partition $\lambda/\mu$.
Let $\epsilon_2$ ($\epsilon_2'$) denote the box bellow (on the left of)
$\epsilon_1$ ($\epsilon_1'$), if there is one, and  the box on the
left of (above) $\epsilon_1$ ($\epsilon_1'$) otherwise. 
Define by recurrence the balanced pairs $\epsilon_i,\epsilon_i'$ until
$c(\epsilon_l)=-N/2+1, c(\epsilon_l')=-N/2$ if $N$ is even or
$\epsilon_l=\epsilon_l', c(\epsilon_l)=(1-N)/2$ if $N$ is odd.
By construction, the set of boxes $\{\epsilon_i,\epsilon_i'\}^l_1$ belongs
to a balanced removable border strip of width one or, simply, a \emph{balanced strip}.
One can repeat the same reasoning with
the Young tableau of $\lambda/\{\epsilon_i,\epsilon_i'\}^l_1$ (which is not necessarily in
the same block as $\mu$ because of ii)).

Thus, we clearly see that partitions $\lambda$ in the same block can be constructed
by dressing up with balanced strips a certain partition $\mu$ with no
removable balanced strips.
Denote by $\eta$ the  balanced strip of smallest length addable to $\mu$.
If $N$ is odd denote by $\bar{\mu}$ the minimal partition $\bar{\mu}/\mu=\eta$.
If $N$ is even  denote by $\bar{\mu}$ the minimal partition
$\bar{\mu}/\mu=\eta$ only if the two boxes with content $-N/2,1-N/2$ in $\eta$ are
disposed horizontally and  $\bar{\mu}=\mu$  otherwise.
Partitions which are of the form $\mu$
dressed up with an even (odd) number of balanced strips are in the same
block as $\mu \, (\bar{\mu})$.
Note that it is irrelevant in what order the strips are dressed on
$\mu$. Also,  there cannot be two balanced strips of the same length.
Thus, a partition $\lambda$ in the block $\mu\, (\bar{\mu})$
is unambiguously  specified by the length of balanced
strips in the skew partition $\lambda/\mu$.

We claim now and show bellow that a block of
$\osp(R|2S)$ is composed, in the partition notation of
sec.~\ref{sec:superalg}, of hook shaped partitions
built up by dressing with balanced strips an atypical  partition with no
removable balanced strips.\footnote{This means that condition ii) is
  relaxed when partitions are viewed as
  $\osp(R|2S)$ weights. In particular the minimal partitions
  $\mu,\bar{\mu}$ discussed above are in the same block of  $\osp(R|2S)$ if there are
  hook shaped and atypical. Condition ii) is clearly related to the
  $\mathbb{Z}_2$ we neglect by looking at the representation theory of
  $\osp(R|2S)$ instead of $\OSp(R|2S)$.}
For that we need to reformulate the original block result \cite{Serganova98} for $\osp(R|2S)$.

Let the \emph{degree of atypicality}
$k$ of a dominant weight $\Lambda$,
be the dimension of the subspace $\mathcal{A}$
of the root lattice orthogonal to $\Lambda +\rho$, where $\rho$ is the 
Weyl vector of $\osp(R|2S)$.
Each atypicality condition in eq.~(\ref{e:1_atyp},\ref{e:2_atyp}) is,
in fact, an orthogonality condition between an odd root $\delta_i\pm
\epsilon_j,\, \epsilon_j\neq 0$ and $\Lambda+\rho$.
Therefore, $k$ is the number of odd roots orthogonal to each other and
to $\Lambda+\rho$ or, equivalently, the number of atypicality
conditions labeled by couples $(i,j)$ with distinct $i$ and $j$.
From the definition of the highest weight module $V(\Lambda)$ it is clear that 
irreducible finite dimensional components of $V(\Lambda)$
must have dominant weights
of the form $\Lambda - \sum \mathbb{N} \alpha$, where the sum is over
all odd positive roots $\alpha$ spanning $\mathcal{A}$.

Consider a $\osp(R|2S)$ weight $\lambda$, which, in the notation of
app.~\ref{sec:superalg}, has symplectic part $\rho$ and orthogonal
part $\sigma$.
Suppose $\rho_{n+1},\rho_{m+1}$ are the first columns of $\lambda$
satisfying $\rho_{n+1}\leq r-S+n$ and  $\rho_{m+1}\leq m+R-S-\rho_S-1$.
Then, one can find rows $i_j$, such that
$\lambda_{i_j}< S$ and the pairs $(i_j,j)$ satisfy
the atypicality condition~\eqref{e:2_atyp} for $m<j< n$ if $R$ is odd
and $m<j\leq n$ if $R$ is even and the atypicality
condition~\eqref{e:1_atyp} for $n\leq j$.
Indeed,  from  eq.~\eqref{e:2_atyp} with $\sigma_i=0$ the condition $m<j$ implies $i_j>
\rho_S$ and thus $\lambda_{i_j}<S$, while  $i_j\leq r$ implies $j\leq n$
if $R$ is even and $j<n$ is $R$ is odd.
From eq.~\eqref{e:1_atyp} with $\sigma_i=0$ the condition $n\leq j$
implies $i_j\leq r$ while $i_j>\rho_S$ follows directly from
$\rho_j\geq \rho_S$.

\begin{figure}
   \psfrag{r}{$r$}  
   \psfrag{S}{$S$}
   \psfrag{n}{$n$}
\psfrag{r-S+n}{$r-S+n$}
   \centerline{\includegraphics{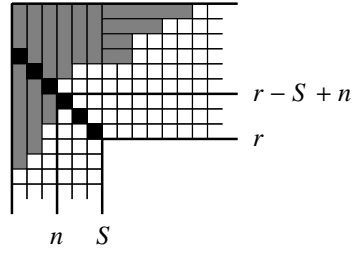}}
   \caption{A partition $\lambda$ drawn on a hook shaped square lattice
     and      fitting exactly inside a hook with foot width $n$ and arm width
     $r-S+n$. The black boxes represent the diagonal of squares
     with content $c=S-r$.}
   \label{fig:imagine}
 \end{figure}

Conversely, if $\rho_j,\,m<j$ satisfies an atypicality condition
with $\sigma_i=0$, then $\lambda_i<S$.
As shown in fig.~\ref{fig:imagine}, $n$ is the width of the foot of the narrowest hook with arm
width $r-S+n$ in which the Young tableau of $\lambda$ can be drawn in.

Two atypicality conditions $(i,j)$ and $(i',j')$ are
called independent if $i\neq i'$ and $j\neq j'$.
Clearly, conditions $(i_j,j)$ are pairwise independent for $m+1\leq
j< n$ and for $n\leq j$.
Let us show that an atypicality condition $(i_j,j),\, m+1\leq j<n$
is independent of conditions  $(i'_{j'},j'),\, n\leq j'$ iff
there is a row shorter then $S$ such that the box $\epsilon$ at the end this row
and the box $\epsilon'_j$ at the end of column $j$ are balanced.

In order to do that it is useful to imagine the partition $\lambda$
drawn on an infinite square lattice, as in fig.~\ref{fig:imagine},
with each square having its content written
inside. The following cases are possible
\begin{itemize}
\item Suppose first that there is no box with at the end of the row $j$.
From $j<n$ follows $j\leq S-r$. Observe that the column $1+S-r \leq
j'=1+S-i_j$ is also empty.
Therefore $i'_{j'}=i_j$ and the atypicality conditions $(i_j,j)$ and
$(i'_{j'},j')$ are not independent.

\item Let $\rho_j\neq 0$ and let $\epsilon_j$ be the box at the end of
column $j$.
No suppose that $\lambda$ has a rightmost box $\omega$ with content
$c=2-N -c(\epsilon'_j)$ and let $j'$ be the column of that box.
Condition $\rho_j\leq m+R-S-\rho_S-1$ gives $c\leq c(\epsilon_S')+1+m-j$.
The equality sign cannot hold because otherwise $j=j'=S$ which
contradicts $j<n$.
Thus $1+S-r \leq c=1+S-i_j<c(\epsilon_S)$ and therefore $S-r\leq n \leq j'<S$.
If $j'-\rho_{j'}<c$ then there is a box $\epsilon$ bellow
$\omega$, which is balanced with $\epsilon'_j$ and has no box to the
right, thus it is the end of a row shorter then $S$.
If  $j'-\rho_{j'} = c$ then comparing the $i$'s from the
two atypicality conditions we get $i_j=i'_{j'}$ and the two atypicality
conditions are not independent.
\item Finally, if there is no box with content $c$ then the column $j'=c$
gives an atypicality condition $(i'_{j'},j')$ with $i_j=i'_{j'}$.
\end{itemize}

Next, by the definition of $m$, a column $j\leq m$
can satisfy an atypicality condition only with a row $\lambda_i\geq S$.
After inserting $\sigma_i=\lambda_i-S$ in
eqs.~(\ref{e:1_atyp},\ref{e:2_atyp}) we get
\begin{align}\label{eq:hooklength}
  &\lambda'_j+\lambda_i+1-i-j=0\\ \label{eq:bcond}
   &(j-\lambda'_j)+(\lambda_i-i)=1-N.
\end{align}
The lhs in eq.~\eqref{eq:hooklength} is the hook length of the box
in the row $i$ and column $j$ of $\lambda$, thus, always positive.
On the other hand, eq.~\eqref{eq:bcond} requires the box $\epsilon'_j$
at the foot of column $j$ be balanced with the box $\epsilon_i$ at the
end of row $i$.

In the end, we see that there are two sources for \emph{independent}
atypicality conditions satisfied by a weight $\lambda$.
First, if $n$ is the width of the foot of the narrowest hook with arm   
width $r-S+n$, in which the Young tableau of $\lambda$ can be drawn
in, then there are $p:= S-n$ atypicality conditions
satisfied by the weight and we call them of type 1.
Second, to each balanced pair of boxes $\epsilon,\epsilon$, such that
$\epsilon$ is a box at the end a row and $\epsilon$ is
a box at the end of a column, corresponds an atypicality
condition of type 2.
If $\lambda$ satisfies $q$ atypicality condition of type 2  then the
degree of atypicality of the weight is $k=p+q$.

Let $\{\epsilon_{i_l},\epsilon'_{j_l}\}^{q}_1,\, j_1<,\dots,<j_{q}$
be the set of balanced pairs satisfying atypicality conditions of
type 2.
Applying the iterative construction explained above
to  the boxes $\epsilon_{i_q},\epsilon_{j_q}'$ one can see that there is a  removable
balanced strip $\eta_{j_q}$ in $\lambda_{q}:= \lambda$, with its ends in
$\epsilon_{j_q},\epsilon_{j_q}'$.
Clearly, by the same reasoning, one can identify a new balanced
strip $\eta_{q-1}$ removable in
$\lambda_{q-1}:=\lambda_{q}/\eta_{q}$.
The end $\lambda_0$ of this iterative procedure has no more removable
balanced strips.
Note that $\lambda_0$ satisfies $k$
atypicality conditions all of type $1$ and the sequence of weights
$\lambda_0,\dots,\lambda_q$ has the same degree of atypicality $k$.

In order to complete the proof of
the claim it remains to notice two things.
First, if $\alpha^1_i$ is an odd
root generating an atypicality condition of type 1, then $\Lambda -
\sum_{i=1}^p \mathbb{N} \alpha^1_i$ is not dominant.
Second, if $\alpha^2=\delta_j+\epsilon_j$ is an odd root generating an atypicality
condition of type 2, then $\Lambda$ has a removable strip with its
ends in the last box $\epsilon_i$  of row $i$ and $\epsilon_j'$ of
column $j$ and $\Lambda - \alpha^2_i$ is dominant and can be represented by a
partition of the form $\lambda/\{\epsilon_i,\epsilon_j'\}$, where
$\lambda$ is the Young tableau of $\Lambda$.
 
\section{Modification rules and $\OSp(R|2S)$ associate weights}
\label{sec:mod_rules}

The explicit form of the characters of classical groups is easier
derived in the limit of infinite rank of the corresponding
Lie algebra.
The inverse limit exists and is given by the \emph{modification rules}
for characters.
The concepts of infinite rank and inverse limit are rigorously defined
for the case of Schur symmetric functions,
connected to the irreducible characters of $\GL(N)$, in
\cite{MacDonald_book}.
Let us clarify this point.

The characters of classical groups, evaluated on a group element,
are polynomials in the eigenvalues of that element in the defining
representation for the group.
The infinite rank limit corresponds to considering polynomials depending
on an infinite number of such variables.
Irreducible characters are polynomials with a very specific symmetry,
which is not obscured by the restriction of finite number of
variables in the infinite rank limit.\footnote{For instance, in the case of Schur
  symmetric functions this specific symmetry is  the  Littlewood-Richardson rule.}
These objects are known as \emph{symmetric functions}.
When the number of variables is set finite most symmetric functions
become functionally dependent.
Once an algebraically independent subset of symmetric functions is
chosen, which is the actual set of characters in the case of classical groups,
the modification rules ``for characters'' represent arbitrary symmetric functions
along this basis.

One can introduce generalized symmetric functions $sc_\mu$ for the
supergroup $\OSp(R|2S)$ according to eqs.~(\ref{eq:elchar}, \ref{eq:onchar}),
see \cite{Bars81}, \cite{Ram98}.
The major difference with respect to classical groups is that
functionally independent generalized
symmetric functions are no longer irreducible characters of the supergroup.
However, modification rules for $sc_\mu$ exist
and have been derived in \cite{King87}.
We bring them bellow in the form of eq.~\eqref{eq:mod_r} with the notations of our paper.

Suppose that $\lambda$ is a typical $\osp(R|2S)$ weight.
Then, according to \cite{King87}, only $sc_\mu$ with  $\mu$ of the form
$\lambda$ dressed by balanced strips $\eta_1, \dots,\eta_m$ modify
to $sc_\lambda$
\begin{equation}
  \label{eq:modtyp}
  sc_\mu = \varepsilon^m w(\mu/\lambda) sc_\lambda,
\end{equation}
here $\varepsilon$ is the superdeterminant representation.
We have also introduced the weight function $w(\mu/\lambda)=
\prod_{i=1}^m(-1)^{c_i-1}$ defined on skew partitions composed of balanced
strips and  $c_i$ is the number of columns
in $\eta_i$.

Consider now, the $\osp(R|2S)$ block labeled by the weight $\nu$ with degree of
atypicality $k$ and no removable balanced strips.
Then, any weight $\lambda$ in the block $\mathsf{B}_\mu$ of $\mu$ is of the form $\nu$
dressed up by $q_\lambda\leq k$ balanced strips.
Then, according to \cite{King87}, $sc_\mu$ with $\mu$ of the
form $\nu$ dressed up by  $m\geq k+1$ balanced strips modifies to
\begin{equation}
  \label{eq:modatyp}
  sc_\mu = \sum_{\lambda\in \mathsf{B}_\nu} C_{m- k-1}^{m-q_\lambda-1} w(\mu/\lambda)
  (-1)^{k-q_\lambda}\varepsilon^{m-q_\lambda}sc_\lambda.
\end{equation}

%For an $\OSp(R|2S)$ associate weight $\lambda^*\equiv\varepsilon\times
%\lambda$ define $sc_{\lambda^*}=\varepsilon sc_\lambda$.

For a typical weight $\lambda$ we put $\lambda^*$ equal
to $\lambda$ dressed up by the balanced strip of minimal length $\eta_1$ if
$\lambda'_S<r$.
In order to prove that  $sc_{\lambda^*}=\varepsilon sc_\lambda$ one
has to show that $\eta_1$ runs over an odd number
of columns.

%and $\lambda$ if $\lambda_S'=r$.

%One can proove that $w(\lambda^*/\lambda)=1$ and, thus, $w(\mu/\lambda)=w(\mu/\lambda^*)$.

Let us prove that $\eta_1$ runs over an odd number 
of columns $c_1$ if $\lambda'_S<r$ and an odd (even) number of columns if
$\lambda'_S=r$ and $R$ is odd (even).

\begin{figure}
  \psfrag{0}{$\epsilon_1$}
 \psfrag{1}{$\epsilon_2$}
 \psfrag{a}{$a$}
 \psfrag{b}{$b$}
 \psfrag{c}{$c$}
 \psfrag{d}{$d$}
 \centerline{\includegraphics{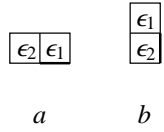}}
 \caption{Configuration of lowest boxes $\epsilon_1,\epsilon_2$
   in $\lambda$ with content $1-N/2,-N/2$ when $N$ is even.
   Thick lines represent the border of the Young tableau of
   $\lambda$.}
 \label{fig:corner}
\end{figure}

Indeed, each box in $\eta_1$ belongs either to a horizontal or a vertical
part  of the strip, except for the boxes at the corners of $\eta_1$, 
which belong to both.
We say the balanced pair $\epsilon,\epsilon'\in \eta_1$ has an allowable
configuration if both boxes belong either to horizontal or vertical 
parts of $\eta_1$, otherwise $\epsilon,\epsilon'$ has a non
allowable configuration.
As discussed in app.~\ref{sec:tmb}, to every non allowable configurations
of  $\epsilon,\epsilon'$ with content $c,c'$ corresponds a removable
balanced strip in $\lambda$ with its ends in the border boxes 
with content $c-1,c'+1$ or $c+1,c'-1$ depending on weather $\epsilon$ is on a 
horizontal or a vertical part of $\eta$.
Because $\lambda$ is a $\osp(R|2S)$ typical weight and, thus, has no
removable balanced strips,
there are only allowable configuration of balanced pairs in $\eta_1$. 
Thus, a balanced pair in $\eta_1$, which is not in the same
column, indexes either two different columns or none. 
There is at most one  column containing the whole balanced pair  and it appears 
always if $N$ is odd and only for $\lambda'_S<r$ if $N$ is even,
as shown in fig.~\ref{fig:corner}. 
Thus, $c_1$ is always odd for $R$ odd and even only if $\lambda'_S=r$
for $R$ even.

Let $\eta_i$ denote the $i$th lowest length strip addable to $\lambda$
and $c_i$ the number of columns in it.
One can prove by the same method that
$\eta_{i+1}/\eta_i$ has an even number of columns, one of
which is already in $\eta_i$ and, consequently, $c_{i+1}-c_i$ is odd.

Again by the same method it is possible to prove that $\eta_1+\eta_2$ has an
even number of columns.
This is because $\eta_2$ contains a substrip $\eta'_1$ which can be
obtained by moving down along the diagonal the strip $\eta_1$.
Applying what was said above about $\eta_{i+1}/\eta_i$  to
$\eta_2/\eta_1'$ we see that $w(\eta_1+\eta_2)=-1$.  

If $\lambda$ is typical and $\lambda_S'<r$ then $\mu=\lambda+\eta_1+\eta_2$
is the next partition in the block of $\lambda$, while if $\lambda_S'=r$ then
$\mu=\lambda+\eta_1$
is the next partition in the bloc of $\lambda$.
Therefore we have just shown, as claimed in sec.~\ref{heart} that
$sc_\mu=-sc_\lambda$.

Thus, the two cases in eq.~\eqref{eq:modtyp} corresponding to the
parity of $m$ can be simply written as $sc_\mu=w(\mu/\lambda)sc_\lambda$ if
$\lambda\preceq \mu$ and  $sc_\mu=w(\mu/\lambda^*)sc_{\lambda*}$ if
$\lambda^*\preceq \mu$ because $w(\mu/\lambda)=w(\mu/\lambda^*)$.

We do not now how to explicitly define the associates of atypical
weights for general $\osp(R|2S)$. 

%For singly atypical $\osp(R|2S)$ weights $\lambda$ with $\lambda_S'<r$
%we define the associate partitions a following:
%if $\lambda$ has no removable balanced strips then
%$\lambda^*/\lambda=\{\eta_1,\eta_2,\eta_3\}$,
%if $\lambda$ has one removable balanced strip $\eta_1$ then
%$\lambda^*/\lambda=\{\eta_2,\eta_3\}$, otherwise
%$\lambda^*/\lambda=\eta_1$.

\end{appendix}
\end{document}